\documentclass[prb,floats,twocolumn,floatfix]{revtex4}
\usepackage{natbib}
\usepackage{graphicx}
\usepackage{color}
\usepackage{ulem}
\definecolor{britishracinggreen}{rgb}{0.0, 0.26, 0.15}
\newcommand{\bq}{{\bf q}}
\newcommand{\bk}{{\bf k}}

\newcommand{\bQ}{{\bf Q}}
\newcommand{\reff}[1]{{(\ref{#1})}}

\newcommand{\ed}{\epsilon_d}
\newcommand{\ep}{\epsilon_p}
\newcommand{\ex}{\epsilon_x}
\newcommand{\ey}{\epsilon_y}
\newcommand{\tpp}{t_{pp}}
\newcommand{\tpd}{t_{pd}}

\newcommand{\vpp}{V_{pp}}
\newcommand{\vpd}{V_{pd}}

\begin{document}
\title{Instability towards Staggered Loop Currents in the Three-Orbital Model 
for Cuprate Superconductors}
\date{\today}
\author{S. Bulut}
\affiliation{Theoretical Physics III,
Center for Electronic Correlations and Magnetism, Institute of Physics,
University of Augsburg, 86135 Augsburg, Germany} 
\author{A. P. Kampf\,}
\affiliation{Theoretical Physics III,
Center for Electronic Correlations and Magnetism, Institute of Physics,
University of Augsburg, 86135 Augsburg, Germany}
\author{W. A. Atkinson}
\affiliation{Department of Physics and Astronomy, Trent
  University, Peterborough Ontario, Canada, K9J7B8}
\begin{abstract}
We present evidence for the existence of a spontaneous instability towards an 
orbital loop-current phase in a multiorbital Hubbard model for the CuO$_2$ 
planes in cuprates. Contrary to the previously proposed $\theta_{II}$ phase 
with intra-unit cell currents, the identified instability is towards a 
staggered pattern of intertwined current loops. The orbitally resolved current
pattern thereby shares its staggered character with the proposal of d-density 
wave order. The current pattern will cause a Fermi surface reconstruction and 
the opening of a pseudogap. We argue that the pseudogap phase with 
time-reversal symmetry breaking currents is susceptible to further phase 
transitions and therefore offers a route to account for axial incommensurate 
charge order and a polar Kerr effect in underdoped cuprates.
\end{abstract}
\maketitle

\section{Introduction}

There is now considerable evidence in underdoped cuprate high-temperature 
superconductors for a cascade of phase transitions, starting at high 
temperatures with the pseudogap onset at $T^\ast$, followed by incommensurate 
charge order (ICO) at $T_\mathrm{co}<T^\ast$ and superconductivity at 
$T_c < T_\mathrm{co}$. In addition, broken time-reversal symmetry has been
associated with $T^\ast$ and a Kerr rotation is measured below a temperature 
$T_\mathrm{Kerr}$ with $T_\mathrm{Kerr}\sim 0.75 T^\ast$ over wide doping 
range. At present, there is no unifying theory that explains this intriguing 
sequence of transitions.

Ultrasound spectroscopy suggests that $T^\ast$ corresponds to a true 
thermodynamic phase transition \cite{Shekhter:2013eh} ; this finding 
challenges the viewpoint that the pseudogap arises as a correlation induced 
phenomenon in a symmetry unbroken paramagnetic phase. Below $T^\ast$ 
spin-polarized neutron scattering experiments detected weak magnetic moments 
\cite{Fauque2006,Li2008,Li2011,Sidis2013}. These moments appear to preserve the 
translational symmetry of the lattice, and led to the proposal of intra-unit 
cell loop currents (LCs) \cite{Varma2006}. However, the so-called $\theta_{II}$ 
LC phase by itself has difficulty explaining the partial gapping of charge 
excitations \cite{Varma:2014id}. While variational methods favored the 
existence of LC phases in finite clusters \cite{Weber2009,Weber2014}, 
alternative numerically exact analyses reported no evidence for the 
$\theta_{II}$ phase \cite{Thomale2008,Kung2014}.

The Kerr effect \cite{Xia2008,RHe2011} that sets in below $T_\mathrm{Kerr}$ is 
further evidence for time-reversal symmetry breaking, but also requires that 
mirror symmetries be broken \cite{Wang2014}. Throughout much of the cuprate 
phase diagram, $T_\mathrm{Kerr}$ and $T_\mathrm{co}$ \cite{Wu2011,Ghiringelli2012,Chang2012vf,Neto2014,Comin2014,Kohsaka2007,Wise2008,Lawler2010,Wu2015} are 
close, which has motivated further proposals in which fluctuating charge- 
\cite{Wang:2014fr,Wang2014} or pair-density wave states 
\cite{Lee:2014ka,Agterberg:2015bo} generate spontaneous current patterns with 
broken mirror symmetries. These scenarios assume a heirarchy of transitions 
associated with distinct symmetry breakings needed to form the fully ordered 
density-wave state. Other proposals follow the common theme that the pseudogap 
results from the competition between two or more order parameters 
\cite{Efetov2013,Pepin:2014tb,Chowdhury2014,Hayward:2014eo}.

The ICO phase involves predominantly a charge transfer between oxygen orbitals 
in the CuO$_2$ planes 
\cite{Kohsaka2007,Wise2008,Lawler2010,Achkar2015,Comin2015}. This challenges 
notions of immutable CuO$_2$ bands, and points to the necessity to employ 
multiorbital models for the ICO phase 
\cite{Fischer2011,Bulut2013,Atkinson2014}. Here, we support this reasoning 
and show that orbital resolved intra-unit cell physics is important throughout 
the pseudogap regime.

\begin{figure}
	\includegraphics[width=5cm]{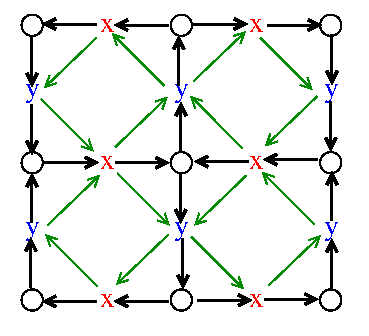}\vspace{-0.5cm}
	\caption{(Color online) Staggered pattern of spontaneous loop currents.
        Open circle,``x" and ``y" symbols denote Cu$d_{x^2-y^2}$, O$p_x$ and 
        O$p_y$ orbitals, respectively. Currents along $p$-$d$ bonds (black 
        arrows) are about three times stronger than those in $p$-$p$ bonds 
        (green arrows).}\vspace{-0.5cm}
	\label{f:pipi_cartoon}
\end{figure}

In this work we report the results of an unbiased calculation for a 
three-band model of CuO$_2$ planes which verifies the existence of an 
instability towards a staggered pattern of intertwined LCs 
(Fig.\ \ref{f:pipi_cartoon}). This ``$\pi$LC'' phase is different from the 
anticipated $\theta_{II}$ phase, but shares its ordering wavevector 
${\bf Q}=(\pi,\pi)$ with the earlier phenomenological proposal of LCs in the
$d$-density wave (DDW) state \cite{Chakravarty2001,Chakravarty2004}. In the 
$\pi$LC phase the Fermi surface reconstructs to form hole pockets with 
a concomitant pseudogap-like structure in the electronic spectrum. 
For realistic parameters, the ICO reported previously for the same model 
Hamiltonian \cite{Bulut2013} is subleading to the $\pi$LC instability. Yet, the
presence of staggered order favors a subsequent instability towards ICO with 
axial wavevectors connecting the tips of the hole pockets \cite{Atkinson2014}, 
consistent with experiments. The charge modulation of the latter necessarily 
breaks mirror symmetries and will hence allow for a polar Kerr signal
\cite{Wang2014}.  
This scenario is offered as a proposal for the cascade of phase transitions in 
the pseudogap regime of underdoped cuprates.


\section{Hamiltonian}
\begin{figure}
    \includegraphics[width=6cm]{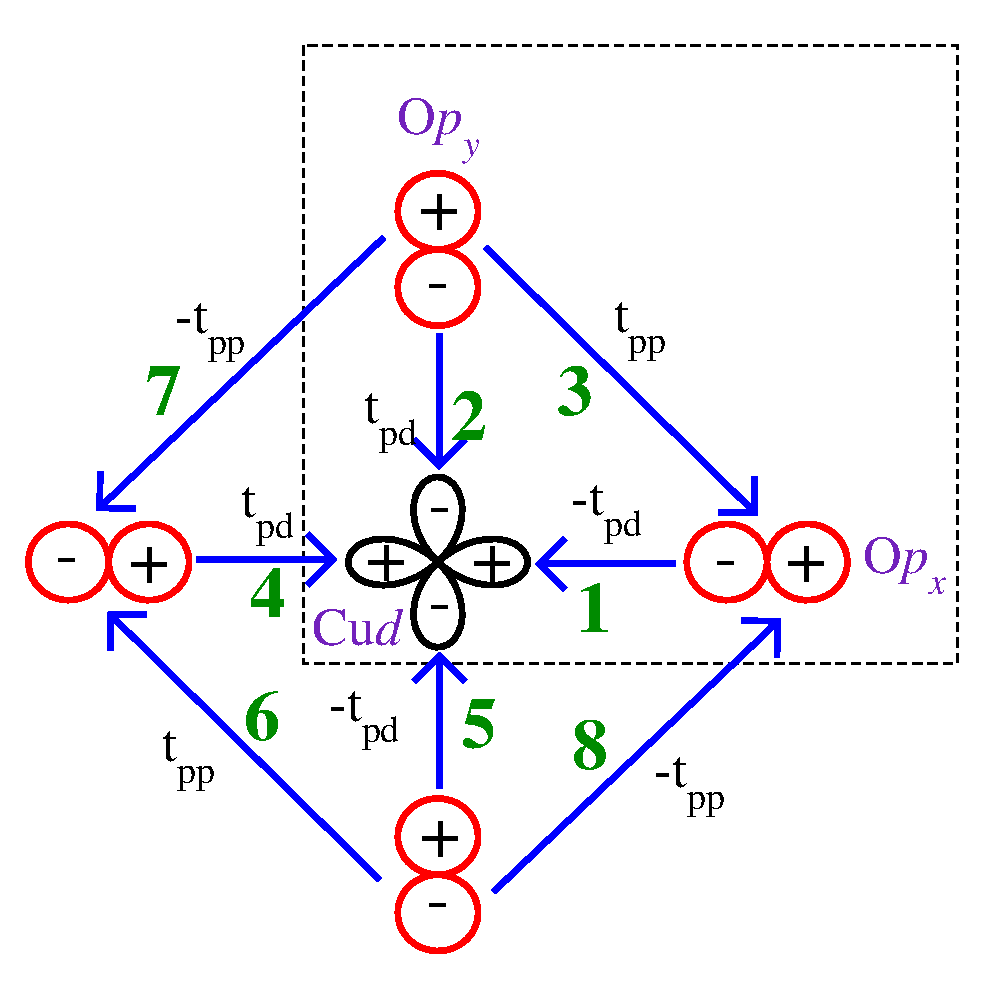}
    \caption{(Color online) Unit cell of a CuO$_2$ plane (dashed box). The 
    orbital phase convention is indicated by the sign of the hopping matrix 
    elements. Numbers in green enumerate the inequivalent bonds, and the 
    direction of the blue arrows indicates a positive sign of the current flow 
    for the current operator definitions given in Appendix~\ref{s:jops}.}
    \label{f:unitcell}
\end{figure}

The unit cell of a single CuO$_2$ plane is shown in Fig.~\ref{f:unitcell}, along with the choice of orbital phases and the corresponding signs of the hopping terms.

The non-interacting part of the three band model is given by
\begin{equation}
\hat H_0 
    = \sum_{i\alpha\sigma} \epsilon_{i\alpha} \hat n_{i\alpha\sigma}
    + \sum_{i\alpha j\beta \sigma} t_{i\alpha j\beta}\hat c^\dagger_{i\alpha\sigma} \hat c^{}_{j\beta\sigma} \label{e:h0real}
\end{equation}
where $i$ and $j$ are unit cell labels, $\alpha$ and $\beta$ are orbital labels, $\sigma$ is the spin label, $\epsilon_{i\alpha}$ is the orbital energy, $\hat n_{i\alpha \sigma}$ is the number operator, $t_{i\alpha j\beta}$ is the tunneling matrix element between orbital $i\alpha$ and $j\beta$, and $\hat c_{i\alpha\sigma}$ and $\hat c^\dagger_{i\alpha\sigma}$ are annihilation and creation operators. Below, we suppress the spin labels.

Using the translational invariance, $\hat H_0$ can be Fourier transformed to reciprocal space:
\begin{equation}
\hat H_0 = \sum_{\bk} \Psi^\dagger_{\bk} {\bf H}_0(\bk) \Psi_{\bk} 
\end{equation}
where $\Psi_{\bk}^\dagger = [ \hat c^\dagger_{\bk d },\, \hat c^\dagger_{\bk x},\,\hat c^\dagger_{\bk y} ]$, and $\hat c^\dagger_{\bk \alpha}$ ($\hat c_{\bk \alpha}$) is the creation (annihilation) operator for an electron with crystal momentum $\bk$ and orbital $\alpha$. Explicitly,
\begin{eqnarray}
    \hat c^{}_{\bk \alpha} &=& \frac{1}{\sqrt{N}} \sum_{i} e^{-i\bk\cdot{\bf R}_{i\alpha}}\hat c_{i\alpha}\label{e:canni}\\
    \hat c^{\dagger}_{\bk \alpha} &=& \frac{1}{\sqrt{N}} \sum_{i} e^{i\bk\cdot{\bf R}_{i\alpha}}\hat c^{\dagger}_{i\alpha}\label{e:crea}
\end{eqnarray} 
where $N$ is the number of unit cells in the system, and ${\bf R}_{i\alpha}$ is the position vector of $\alpha$'th orbital in $i$'th unit cell. $H_0(\bk)$ is readily obtained by plugging Eqs. \reff{e:canni} and \reff{e:crea} into Eq. \reff{e:h0real}:
\begin{equation}
    {\bf H}_{0}(\bk)
    = \left (
        \begin{array}{ccc}
            \ed & -2i\tpd s_x & 2i\tpd s_y \\
             2i\tpd s_x   & \ex & 4\tpp s_x s_y \\
             -2i\tpd   & 4\tpp s_x s_y    & \ey
        \end{array}
      \right ) 
\end{equation}
where $s_x=\sin(k_x/2)$ and $s_y = \sin(k_y/2)$.
A more convenient form of ${\bf H}_0$ is obtained after the following gauge transformation: 
\begin{eqnarray}
\hat c^{}_{\bk x}			&\rightarrow & i \hat c^{}_{\bk x} \label{e:gtks1}\\
\hat c^{\dagger}_{\bk x}	&\rightarrow & -i \hat c^{\dagger}_{\bk x} \\
\hat c^{}_{\bk y}			&\rightarrow & i  \hat c^{}_{\bk y} \\
\hat c^{\dagger}_{\bk y}	&\rightarrow & -i \hat c^{}_{\bk y}\label{e:gtks4}.
\end{eqnarray}
Hence, the final form of ${\bf H}_0$ is obtained:
\begin{equation}
   {\bf H}_0(\bk) =
    \left (
        \begin{array}{ccc}
            \ed & 2\tpd s_x & -2\tpd s_y \\
			2\tpd s_x	& \ep	& 4\tpp s_x s_y \\
			-2\tpd s_y	& 4\tpp s_x s_y	& \ep		
        \end{array}
    \right ). 
\end{equation}
We set $t_{pd} = 1$ so that it defines the unit of energy.

The interacting part of the Hamiltonian includes the intra- ($U_\alpha$) and inter-orbital ($V_{i\alpha,j\beta}$) Coulomb interactions
\begin{equation}
    \hat H'= \sum_{i\alpha\sigma, j\beta\sigma'} \left [ \delta_{i\alpha, j\beta}(1-\delta_{\sigma,\sigma^\prime} ) U_\alpha + \frac{V_{i\alpha,j\beta}}{2} \right ]\hat n_{i\alpha\sigma} \hat n_{j\beta\sigma^\prime},
\end{equation}
where $i,j$ are unit cell indices, $\alpha, \beta$ are orbital labels and $V_{i\alpha,j\beta}$ is nonzero for nearest-neighbors only.
Throughout, we suppress the spin index $\sigma$, set $\tpp=-0.5$, 
$\ed -\ep = 2.5$,  $U_{d}=9$, $U_p=3$, $\vpd=2.2$, and $\vpp=1$ unless otherwise stated. 
The Hamiltonian $\hat H=\hat H_0 + \hat H^\prime$ is thus the conventional three-band model of cuprates\cite{Emery1987} with a typical parameter set \cite{Hybertsen1989}. 

\section{Interacting current susceptibility}
\label{s:intcursus}

The current operator associated with the bond between sites $i\alpha$ and $j\beta$ is $\hat J_{i\alpha,j\beta}=-it_{i\alpha,j\beta}(\hat c_{i\alpha}^\dagger c_{j\beta}^{} - \hat c_{j\beta}^\dagger \hat c^{}_{i\alpha})$
where $t_{i\alpha, j\beta}$ is the corresponding hopping matrix element. If $\langle \hat J_{i\alpha, j\beta}\rangle > 0$ then current flows from $j\beta$ to $i\alpha$. In momentum space, the current operator along bond $m$ is given by 
\begin{equation}
\hat J_m(\bq)=-i\sum_{\bk}[h_{\alpha\beta}^{m}(\bk,\bq)\hat c^\dagger_{\bk\alpha}\hat c_{\bk+\bq\beta}^{}-h_{\beta\alpha}^{m}(\bk,\bq)\hat c^\dagger_{\bk\beta}\hat c^{}_{\bk+\bq\alpha}]
\end{equation}
 where $\alpha,\beta$ are the orbitals associated with the bond and the matrix elements of the current operators $h_{\alpha\beta}^{m}(\bk,\bq)$ are listed in Table \ref{t:jfactors}. As shown in Fig. \ref{f:unitcell}, there are eight distinct bonds on the CuO$_2$ lattice. Accordingly 
the current susceptibility,
\begin{equation}
\chi^J_{mn}(\bq,i\omega_\ell) = 
        \int_0^\beta d\tau\,e^{i\omega_\ell\tau}
        \langle \hat J_m(\bq,\tau) \hat J_n(-\bq,0)\rangle,
\end{equation}
is an 8$\times$8 matrix, where $\omega_\ell=2\pi\ell T$ denotes the bosonic Matsubara frequencies.
Each matrix element can be decomposed as
\begin{equation}
\chi^J_{mn} = 
\chi^{mn}_{\alpha\beta\alpha'\beta'}
-\chi^{mn}_{\alpha\beta\beta'\alpha'}
-\chi^{mn}_{\beta\alpha\alpha'\beta'}
+\chi^{mn}_{\beta\alpha\beta'\alpha'}
\end{equation}
where
\begin{eqnarray}
\chi^{mn}_{\theta\theta'\gamma\gamma'}(\bq,i\omega_\ell)&=&
\frac{-1}{N}
\sum_{\bk\bk'}h^{m}_{\theta\theta'}(\bk,\bq)h^{n}_{\gamma\gamma'}(\bk',-\bq) \nonumber \\
&& \times   \int_0^\beta d\tau\, e^{i\omega_\ell\tau} \langle 
\hat c^\dagger_{\bk\theta}(\tau) c^{}_{\bk+\bq\theta'}(\tau) \nonumber \\
&& \times
  \hat c^\dagger_{\bk' \gamma}(0) c^{}_{\bk'-\bq\gamma'}(0)  
\rangle. \label{e:intsus}
\end{eqnarray}

\begin{figure}
	\includegraphics[width=\columnwidth]{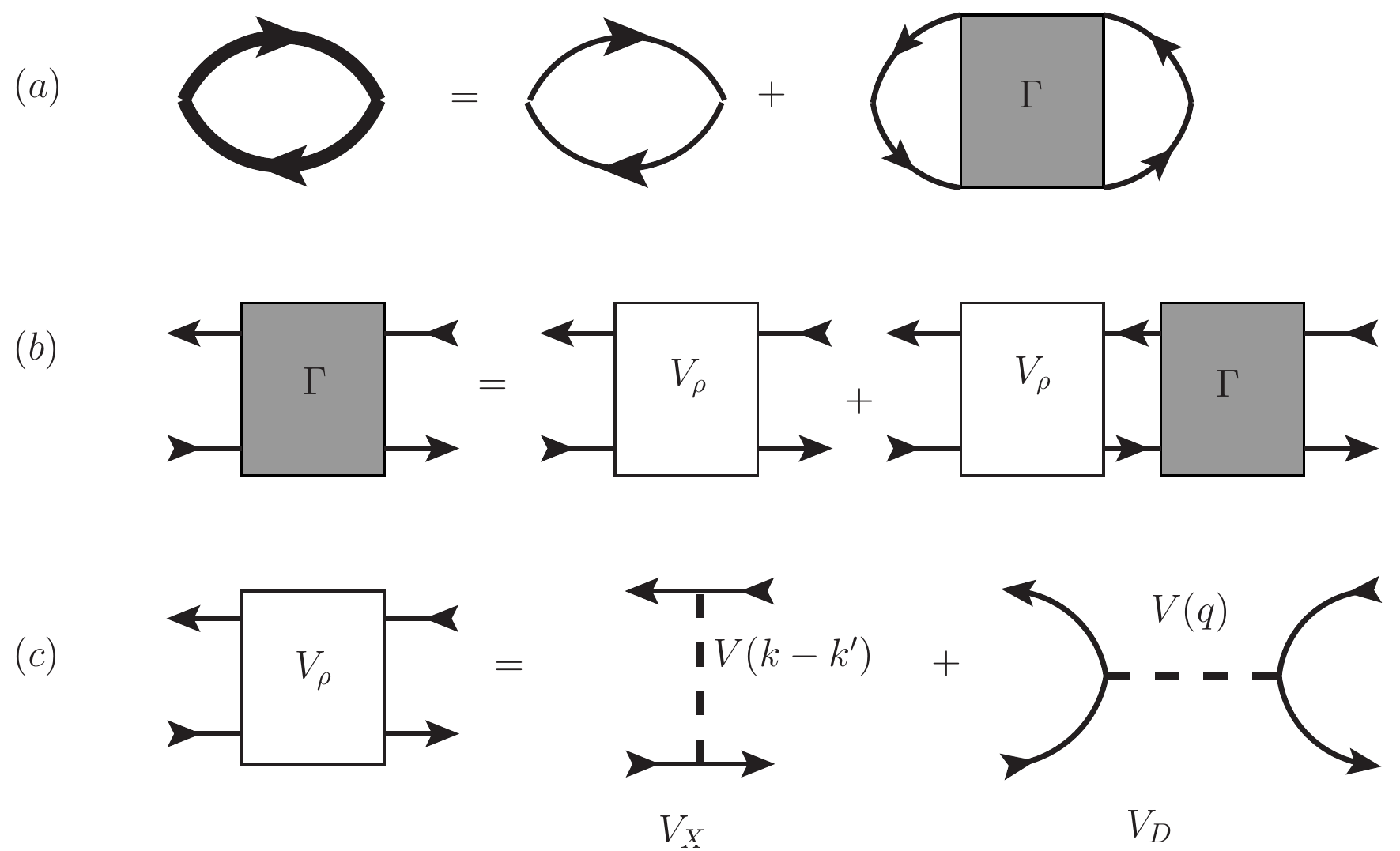}
	\caption{Diagrammatic structure of gRPA: (a) Interacting 
                             susceptibility, (b) vertex function, 
                             (c) effective interaction. 
Reprinted with permission from \cite{Bulut2013}. \copyright American Physical Society.
}
	\label{f:diags}
\end{figure}

Previously, we investigated charge instabilities in the same three-band model with 
non-local interactions using a generalized random phase approximation 
(gRPA)\cite{Littlewood1989,Bulut2013}.
While methods like QMC or cluster DMFT are at first glance more desirable as they are designed to handle strong local correlations, they are less accurate in treating non-local interactions, and are also limited in momentum-space resolution.
Although it neglects strong correlation physics, the gRPA has the advantage that it treats local and non-local interactions on the same footing, and is unbiased with respect to wavevector and to the unit cell-resolved current pattern.
Within gRPA, the 2-particle vertex function includes both exchange and direct 
interaction diagrams and hence also generates combinations of both (see Fig.~\ref{f:diags}), while Green functions remain unrenormalized.

Following Ref.~\cite{Bulut2013}, we project the interactions onto a set of 19 basis functions $g^i_{\alpha\beta}(\bk)$
in orbital and momentum space, leading to a $19\times 19$ matrix equation for the effective interaction vertex $\tilde \Gamma^{ij}(q)$, where $q \equiv (\bq,\omega)$, and $i,j$ label the basis functions.  The basis functions and the interaction vertex are the same as in Ref.~\cite{Bulut2013}.  Closing $\tilde \Gamma^{ij}(q)$ on the left and right with current vertex functions $A^{i,\eta m}_{\alpha\alpha^\prime}(q)$ yields the susceptibility
\begin{equation}
\chi^{mn}_{\alpha\alpha'\beta\beta'}(q)
	= \chi^{0,mn}_{\alpha\alpha'\beta\beta'}(q)
        - \sum_{ij} A^{i,Lm}_{\alpha\alpha'}(q)
        {\tilde \Gamma}^{ij}(q) 
        A^{j,Rn}_{\beta\beta'}(q),
        \label{e:currcomp}
\end{equation}
where
\begin{eqnarray}
	\chi^{0,\ell_1\ell_2}_{\alpha\alpha'\beta\beta'}(q)
	\hspace{-1mm}&=&\hspace{-2mm}\frac{1}{N}
		\hspace{-1mm}
		\sum_{\bk\mu\nu}
		\hspace{-1mm}
		h^{\ell_1}_{\alpha\alpha'}(\bk,\bq)
		M^{\alpha'\beta\beta'\alpha}_{\mu\nu\bk q}
		F^{\nu\mu}_{\bk\bq}(\omega)
		h^{\ell_2}_{\beta\beta'}(\bk-\bq,\bq), \nonumber\\
	A^{i,L\ell}_{\alpha\alpha'}
	&=&	\hspace{-1mm}\frac{1}{N}
		\hspace{-1mm}
		\sum_{\bk\mu\nu\theta\theta'}
		\hspace{-1mm}
		h^{\ell}_{\alpha\alpha'}(\bk,\bq)
		M^{\alpha'\theta\theta'\alpha}_{\mu\nu\bk q}
		F^{\nu\mu}_{\bk\bq}(\omega)
		g^i_{\theta\theta'}(\bk) \\
	A^{i,R\ell}_{\alpha\alpha'}\nonumber
	&=&	\hspace{-1mm}\frac{1}{N}
		\hspace{-1mm}
		\sum_{\bk\mu\nu\theta\theta'}
		\hspace{-1mm}
		h^{\ell}_{\alpha\alpha'}(\bk-\bq,\bq)
		M^{\alpha'\theta\theta'\alpha}_{\mu\nu\bk q}
		F^{\nu\mu}_{\bk\bq}(\omega)
		g^i_{\theta\theta'}(\bk) \\
	M^{\gamma'\theta\theta'\gamma}_{\mu\nu\bk q} 
	&=& S_{\gamma'\nu}(\bk)S^*_{\theta\nu}(\bk)S_{\theta'\mu}(\bk+\bq)S^*_{\gamma\mu}(\bk+\bq) \\
	F^{\nu\mu}_{\bk\bq}(\omega) &=& \frac{f(E_{\bk}^\nu)-f(E^\mu_{\bk+\bq})}{\omega + E_{\bk}^\nu-E_{\bk+\bq}^\mu+i\delta},\label{e:Fw}
\end{eqnarray}
$\ell$ denotes bond indices, $h^{\ell}_{\alpha\alpha'}(\bk,\bq)$ are matrix elements of the 
current operators which are explicitly defined in Table \ref{t:jfactors}, 
$S_{\alpha\nu}(\bk)$ is the $\alpha$th element of the $\nu$th eigenvector of 
${\bf H}_0(\bk)$, $S^*_{\alpha\nu}(\bk)$ is its complex conjugate, $E^\nu_\bk$ are the eigenvalues, $f(E)$ is the Fermi function, and $i$ and $\delta$ in Eq.~\ref{e:Fw} are the complex constant and  a small broadening parameter respectively.
  The bare current susceptibility $ \chi^{0,mn}_{\alpha\alpha'\beta\beta'}(q)$ and the functions $A^{i,L\ell}_{\alpha\alpha^\prime}(q)$ and $A^{i,R\ell}_{\alpha\alpha^\prime}(q)$ differ from Ref.~\cite{Bulut2013} as they contain current operators.

\begin{table}[h]
\begin{tabular}{c|c|l}
	$\ell$ & $\theta\theta'$ & $h^{\ell}_{\theta\theta'}(\bk,\bq)$ \\
	\hline
	1 & dx & $\tpd e^{i(q_x+k_x)/2}$ 		\\
	1 & xd & $-\tpd e^{-ik_x/2}$ 			\\
	2 & dy & $-\tpd e^{i(q_y+k_y)/2}$		\\
	2 & yd & $\tpd e^{-ik_y/2}$ 		  	\\
	3 & xy & $-i\tpp e^{i(q_y-k_x+k_y)/2}$ 	\\
	3 & yx & $-i\tpp e^{i(q_x+k_x-k_y)/2}$ 	\\
	4 & dx & $-\tpd e^{-i(q_x+k_x)/2}$  	\\
	4 & xd & $\tpd e^{ik_x/2}$ 				\\
	5 & dy & $\tpd e^{-i(q_y+k_y)/2}$ 		\\
	5 & yd & $-\tpd e^{ik_y/2}$ 			\\
	6 & xy & $-i\tpp e^{-i(q_y+k_x-k_y)/2}$ \\
	6 & yx & $-i\tpp e^{-i(q_x+k_x-k_y)/2}$ \\
	7 & xy & $i\tpp e^{i(q_y+k_x+k_y)/2} $ 	\\
	7 & yx & $i\tpp e^{-i(q_x+k_x+k_y)/2} $	\\
	8 & xy & $i\tpp e^{-i(q_y+k_x+k_y)/2} $	\\
	8 & yx & $i\tpp e^{i(q_x+k_x+k_y)/2} $  \\
\end{tabular}
\caption{Matrix elements of the current operator. $i$ is the imaginary constant. The overall sign of each term results from three factors: the complex constant in the current operator definitions, the sign of the hopping terms, and the gauge transformation.}
\label{t:jfactors}
\end{table}

We search for the existence of spontaneous currents by following the evolution of the current susceptibility upon cooling. The instability is signalled by a divergence of the momentum-resolved susceptibility at zero frequency $\chi^J_{mn}({\bf q},\omega=0)$.

\section{Results}
\begin{figure}
    \centerline{$\chi_{11}^J(\bq,\omega=0)$}
    \includegraphics[width=0.8\columnwidth]{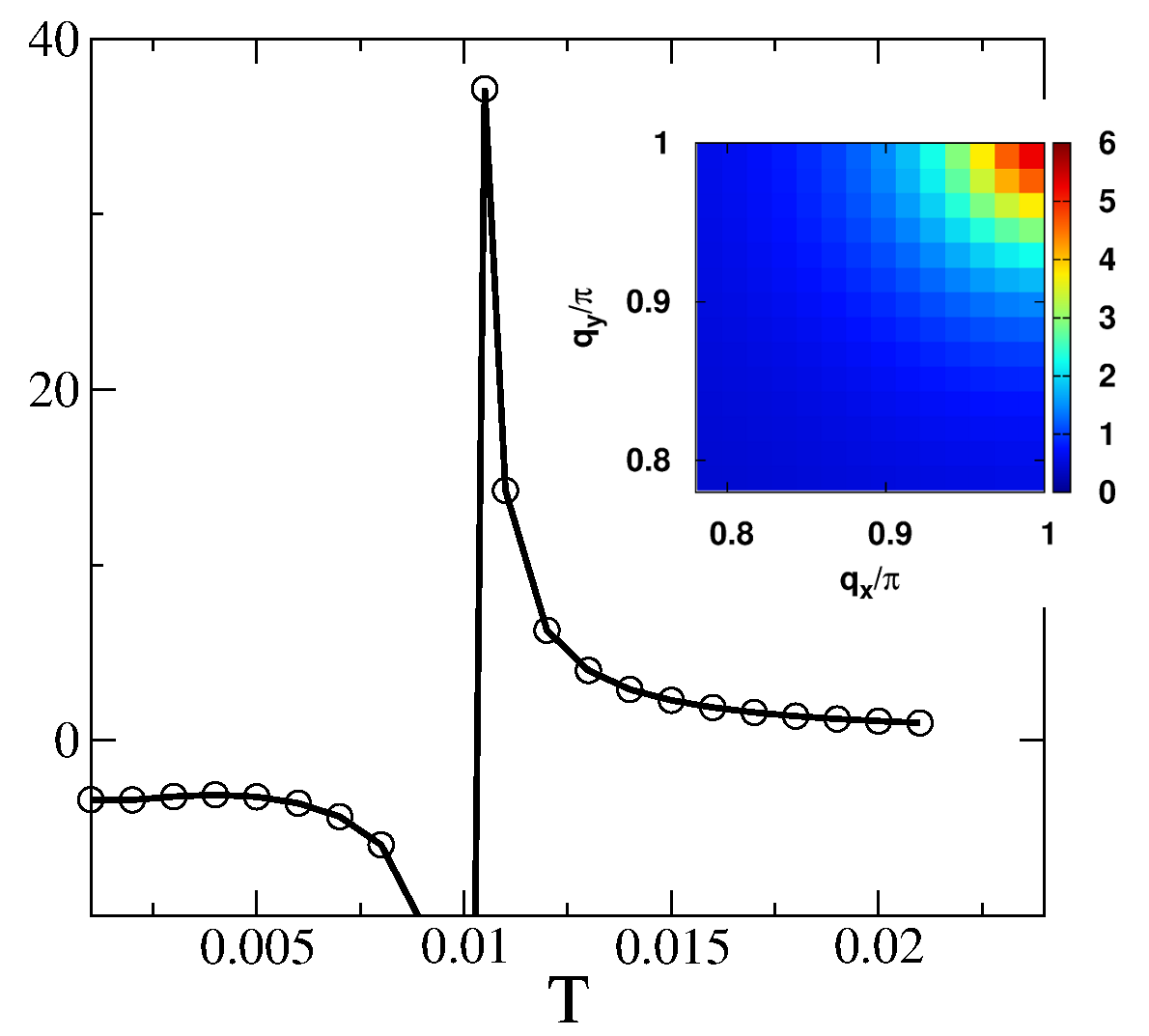}
	\caption{(Color online) Temperature evolution of $\chi_{11}^J(\bQ,\omega=0)$ for $\bQ=(\pi,\pi)$. (inset) $\chi_{11}^J(\bq,\omega=0)$ at $T=0.011$. The parameter values are $\vpd=2.2$, $\vpp=1$, and the hole density is $p=0.13$.}
	\label{f:divergence}
\end{figure}

Figure~\ref{f:divergence} shows typical results for the current susceptibility.
The inset shows that the matrix element $\chi_{11}^J(\bq,\omega=0)$, corresponding to currents along the $d$-$p_x$ bonds, becomes strongly peaked at $\bq=\bQ \equiv (\pi,\pi)$ as the temperature is lowered.  This peak  indeed diverges upon cooling to the critical temperature near $T=0.01$ (main panel), which  signals  an instability towards a current-carrying state.
The ordering wavevector $\bQ$ of the $\pi$LC phase is the same as in the DDW scenario,\cite{Chakravarty2001} and should be contrasted with  the $\theta_{II}$ phase,\cite{Varma2006} for which $\bq={\bf 0}$.

The example result in Fig.~\ref{f:divergence} was obtained for $\vpd=2.2$ and $\vpp=1$. However, the instability towards a $\pi$LC phase persists when $\vpp=0$, and hence is driven by the Coulomb repulsion $\vpd$ between copper and oxygen orbitals.
In fact, also the local interactions $U_d$ and $U_p$ have no effect on the $\pi$LC instability. As we have explicitly verified, the staggered current instability originates from the exchange (ladder only) diagrams.

As previously established\cite{Atkinson2014}, an ICO with a predominant charge redistribution between O$p_x$ and O$p_y$ orbitals can be generated by the Coulomb repulsion $\vpp$ between $Op_x$ and $Op_y$ orbitals. 
Based on the inter-orbital distances, we expect $\vpd > \vpp$, which implies 
that loop currents emerge at higher temperatures than ICO. Indeed, for our parameter 
values, the critical temperature for the $\pi$LC instability is about twice as large as the  critical temperature for ICO.

To determine the bond-resolved $\pi LC$ pattern, we calculate the eigenvector of the leading eigenvalue of the current susceptibility matrix.
In the current operator basis $\hat J_1(\bq),\dots ,\hat J_8(\bq)$, this normalized eigenvector is  $[\,0.48,  -0.48, -0.15,\,0.48,  -0.48,  -0.15,  -0.15,  -0.15\,]$; this eigenvector reveals the direction and the relative magnitudes of the currents on the eight inequivalent bonds: all bonds are involved in the $\pi$LC instability, and for the selected parameter set the currents along the $p$-$d$ bonds are about three times stronger than those along the $p$-$p$ bonds.
The relative strength of the $p$-$d$ and $p$-$p$ currents varies with the ratio $\tpp/\tpd$. 
The wavevector $\bQ$ of the instability further implies that the pattern alternates between adjacent unit cells. We thus obtain the cartoon shown in Fig. \ref{f:pipi_cartoon}, in which two distinct (green and black) and interpenetrating loop currents are evident.
This pattern is similar to the previously proposed current-carrying phases of 
either the DDW -\cite{Chakravarty2001} or the staggered flux-phase type \cite{Hsu1991,Ivanov2000,Tsutsui2002}, but 
differs in having two circulating current loops one of which involves oxygen orbitals only.

The instability towards spontaneous $\pi$LCs  will naturally reconstruct the Fermi surface. 
To explore this  we implement the $\pi$LCs on the mean-field level. The starting point for this calculation is the general identity
 $\hat n_a \hat n_b= ( \hat J_{ab}^{}\hat J^*_{ab}/t_{ab}^2 +\hat n_a+\hat n_b)/2$ 
 that is true for any pair of orbitals $a$ and $b$\cite{He2012b}.
Hence, the non-local density-density interactions $\frac 12 \sum_{i\alpha,j\beta} V_{i\alpha,j\beta} \hat n_{i\alpha} \hat n_{j\beta}$ can be decoupled by introducing the current amplitudes
$z_{i\alpha,j\beta} = \langle \hat J_{i\alpha,j\beta}\rangle=\pm z_{\alpha\beta}$.
The mean-field version of the interorbital Coulomb interactions thus reads:
\begin{equation}
\hat H_{MF}^\prime = {\tilde \epsilon}_p(\hat n_x + \hat n_y) + {\tilde \epsilon}_d \hat n_d \,- \sum_{\langle i\alpha,j\beta\rangle} \frac{V_{i\alpha,j\beta}}{2t_{i\alpha,j\beta}^2}
\hat J_{i\alpha,j\beta} z_{i\alpha,j\beta}
\end{equation}
 where ${\tilde \epsilon}_p= \vpd+2\vpp$ and ${\tilde \epsilon}_d= 2\vpd$ renormalize the orbital energies.   The intraorbital interactions lead to additional Hartree shifts of the orbital energies; these are assumed to be already included in $\epsilon_d$ and $\epsilon_p$.   We obtain the mean field Hamiltonian $\hat H_{MF} = \hat H_0 + \hat H^\prime_{MF}$, with $\hat H_{MF} = \sum_\bk \overline \Psi^\dagger_\bk {\bf H}_{MF}(\bk) \overline \Psi_\bk$,
$\overline \Psi_\bk =  (\Psi_{{\bf k}},\Psi_{{\bf k}+{\bf Q}})^T$, and 
\begin{eqnarray}
{\bf H}_{MF}(\bk)
&=& \left [
        \begin{array}{cc}
                {\bf H}_0(\bk)    & {\bf H}_1(\bk,\bQ) \\
                {\bf H}_1^\dagger(\bk,\bQ) & {\bf H}_0(\bk+\bQ)
        \end{array}
\right ]\, , \\
{\bf H_1}(\bk,\bQ)
&=&  
	\left [
        \begin{array}{ccc}
        0                               & iR_{pd}s_x'   & iR_{pd} s_y' \\
        R_{pd} c_x'   & 0                             & +R_{pp}c_x' s_y'\\
                R_{pd} c_y'   & -R_{pp} s_x' c_y'      & 0 \\
                \end{array}
        \right ]\, ,
\end{eqnarray}
where
$R_{pd}=z_{pd}\vpd/\tpd$, $R_{pp}=2z_{pp}\vpp/\tpp$, $s_x'=\sin[ (k_x+Q_x)/2]$, $c_x'=\cos[ (k_x+Q_x)/2]$, and $s_y',c_y'$ are defined accordingly, and the orbital energies in ${\bf H}_0(\bk)$ are shifted by $\tilde \epsilon_d$ and $\tilde \epsilon_p$.

\begin{figure}
	\includegraphics[width=\columnwidth]{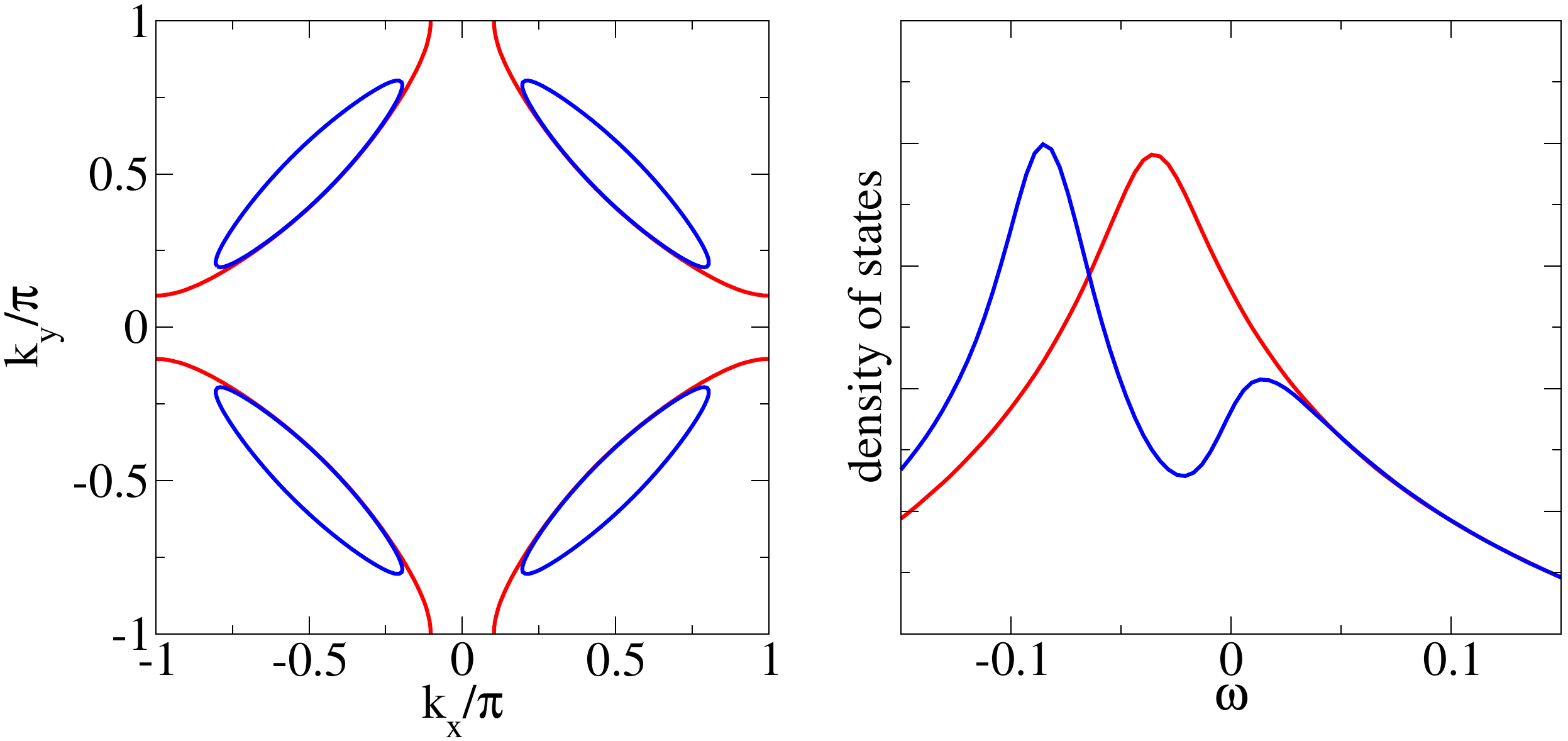}
	\caption{(Color online) Left: Normal (red) and reconstructed (blue) Fermi surfaces. The loop current strengths are set to $z_{pd}=0.04$ and $z_{pp}=z_{pd}/3$. The hole density here is $p=0.10$. Right: Density of states near the Fermi energy.}
	\label{f:fsreconst}
\end{figure}

We show the reconstructed ($z_{pd}=0.04$, $z_{pp}=z_{pd}/3$)
and normal ($z_{pd}=z_{pp}=0$) Fermi surfaces and densities of states in Fig.~\ref{f:fsreconst}.
For very small $z_{pd}$ and $z_{pp}$, hole pockets, as well as small electron pockets near $(\pi,0)$ and symmetry-related points, are formed. 
The shapes of the pockets depend on the strength of the 
loop currents, the hole filling, and the curvature of the unfolded Fermi 
surface. 
With increasing $z_{pd}$ and $z_{pp}$, the electron pockets rapidly disappear, and only the hole pockets persist (Fig.\ \ref{f:fsreconst}).  
If one takes $t_{pd}$ from band structure calculations, $z_{pd}=0.04$ 
corresponds to a $p$-$d$ current of $16$ $\mu$A; empirical bandwidths are $3$ times smaller,\cite{Pasanai2010} giving a current of $5$ $\mu$A and a plaquette
magnetic moment in the range of $0.05\mu_B$ to $0.09\mu_B$ (see Appendix~\ref{s:estim}).

We propose that the pseudogap  appearing at $T^\ast$ arises from the 
$\pi$LC shown in Fig.~\ref{f:pipi_cartoon}. 
Physical properties of staggered currents have been discussed previously
\cite{Chakravarty2001,Laughlin2014prb,Dimov2008,Hsu1991}, and we focus here on aspects 
related to the recent discoveries of charge order and time-reversal symmetry breaking.
The main distinguishing feature of the pseudogap  is the depletion of spectral weight along regions of the Fermi 
surface near $(\pm \pi,0)$ and $(0,\pm \pi)$.
  
This leads to a pseudogap  in the density of states, 
as shown in Fig.~\ref{f:fsreconst} for the $\pi$LC phase. 
In our calculations $z_{pd}$ is about one third of the peak-to-peak pseudogap; experimental pseudogaps of $\sim 100$ meV\cite{Kurosawa2001} therefore suggest 
$z_{pd}\sim0.033$eV, corresponding to a $p$-$d$ current of $\sim 7\mu A$, consistent 
with the estimate above. 

\section{Discussion}
Within the phase with staggered loop currents subsequent phase transtions are 
likely to occur. Notably, we showed previously that persistent discrepancies 
between theory and experiment regarding the ordering wavevector $\bq^\ast$ of 
the charge ordered phase are resolved, if the charge order emerges from a 
preexisting pseudogap phase, rather than causing it. In 
Ref.~\cite{Atkinson2014}, a spin-density wave (SDW) with ordering 
wavevector ${\bf Q}=(\pi,\pi)$ was invoked ad hoc as a proxy for the pseudogap
in underdoped cuprates. While the presence of a static SDW is not supported by 
experiment, we view the $\pi$LC phase instead as a viable alternative phase 
out of which a charge-density wave will form with a $\bq^\ast$ that connects 
adjacent hole pockets.
In the coexistence with charge order, the Fermi surface of the $\pi$LC phase will further reconstruct. The Fermi surface in the coexistence  phase should then serve as the basis to analyze the quantum oscillation experiments which reported evidence for the existence of hole pockets.\cite{Doiron2007,Sebastian2012}

The $\pi$LC phase shares 
neutron-scattering signatures with the DDW state,
specifically an elastic magnetic peak centered at ${\bf Q}=(\pi,\pi)$ and the 
opening of a spin excitation gap. Soon after the original proposal of the DDW 
state it was argued \cite{Chakravarty2001b} that the neutron-scattering data 
obtained by Mook et al. in underdoped YBCO \cite{Mook2001} are consistent with 
the expected features of a DDW state. Other subsequent neutron scattering 
measurements on oxygen ordered ortho-II YBCO instead \cite{Stock2002} reported 
no evidence for the predicted characteristics of an ordered DDW state. The 
conflicting results of these experiments have remained unresolved. 

An obvious signature of the  $\pi$LC phase is that it breaks time-reversal 
symmetry. It does not, however, generate a polar Kerr effect because the 
pattern in Fig. \ref{f:pipi_cartoon} preserves mirror symmetries. Given this 
result, an explanation for the observed nonzero Kerr angle at $T_\mathrm{Kerr}$ 
requires the onset of a further transition that eliminates these 
symmetries.\cite{Wang2014} This could naturally occur, for example, with the 
appearance of incommensurate charge order at $T_\mathrm{co}$. The experimental 
doping dependences of $T_\mathrm{Kerr}$ \cite{Xia2008} and $T_\mathrm{co}$ 
\cite{Huecker2014,Blanco2014} are, however, different; a possible connection 
between the two is therefore not obvious.

It is nevertheless possible that the different symmetries associated with the 
charge 
ordering transition are broken at distinct temperatures.\cite{Wang:2014fr} 
Charge order involves both a continuous broken symmetry associated with the 
spatial lock-in of the charge modulation, and a discrete broken symmetry 
associated with the orientation of $\bq^\ast$.\cite{Nie:2014jj} In this 
scenario, $T_\mathrm{Kerr}$ may signal the onset of a charge-nematic phase, in 
which only the discrete rotational symmetry is broken, while $T_\mathrm{co}$ may 
mark the pinning of the charge modulation by disorder.\cite{Nie:2014jj} 

We are led to trace the origin of the intriguingly complex physics of
underdoped cuprates to the distinct phenomena which emerge from the Coulomb 
interactions in the CuO$_2$ planes: the local Coulomb repulsion on the Cu 
$d$-orbital is the source of antiferromagnetism in the undoped compounds and of
spin-fluctuation mediated $d$-wave superconductivity upon doping. From the 
results of this work we conclude that the non-local interaction $V_{pd}$ can 
cause an orbital current instability, while the non-local interaction $V_{pp}$ 
is responsible for the charge redistribution between O$p_x$ and O$p_y$ orbitals
and incommensurate charge order. 
$V_{pp}$ and $V_{pd}$ weaken the spin-fluctuation mediated pairing 
interaction, which suggests a possible reason why $T_c < T_\mathrm{co}$ 
below optimal doping.
The physics of underdoped cuprates therefore 
appears to reflect the mutual competition and/or coexistence of these ordering 
tendencies.

\section{concluding remarks}

Within gRPA, we find no trace of a $\bq=0$ current instability that has been proposed to exist in the three band model\cite{}. 
Recently, Weber {\it et al.\ } \cite{Weber2014} concluded that the charge-transfer energy, $\epsilon_d-\epsilon_p$, is one of the key parameters for this instability. 
However, over an exhaustive range of this parameter, our calculations did not trace any instability or a sizable enhancement of the $\bq=0$ current susceptibility. 
Intra-unit cell magnetism has been inferred from spin-flip neutron scattering experiments\cite{Li2008,Sidis2013}.
At this stage, our calculations offer no explanation for this observation.

The close connection between axial charge order with a $d$-wave form factor\cite{Achkar2015} and oxygen orbitals makes the three-band model a necessary starting point for the microscopic theory.
In this context, the focus  has so far been on spin fluctuations, and hence on the local Coulomb interactions $U_d$ and $U_p$.
Our findings, however, point to the important role of non-local interactions.
The development of accurate numerical tools capable of handling multi orbital models with intermediate to strong and non-local interactions is, therefore, on demand.

We finally note that, for the typical parameter values used in this work, an SDW phase will, within gRPA, in fact set in at a higher temperature than the $\pi$LC phase. 
This SDW is driven by the local Coulomb interaction $U_d$ on the Cu sites, and it is known that correlation effects beyond the approximations discussed here suppress this SDW. Indeed, the feedback of spin fluctuations renormalizes the electronic structure in a way that reduces the tendency towards antiferromagnetism.  We have neglected this aspect here, since we are primarily interested in the charge degree of freedom in the underdoped part of the cuprate phase diagram where pseudogap and charge order occurs.
Furthermore, we confirmed that the local interactions $U_d$ and $U_p$, the key parameters which control antiferromagnetism and spin fluctuations, do not have any effect on the emergence of the $\pi$LC phase.

It remains yet to be explored how self-energy corrections in our present approach will influence the instability towards spontaneous loop currents.
This is on the agenda for further work on loop currents in the three-orbital model for cuprates.

\section*{Acknowledgements}
A.P.K.\ and S.B.\ were supported by the DFG through TRR80. 
W.A.A.\ acknowledges support by the National Sciences and Engineering Research 
Council (NSERC) of Canada. We thank L.\ Chioncel for helpful discussions.
The calculations were performed on the Linux Cluster of the LRZ in Garching.

\bibliography{loopcurrents_custom}

\begin{thebibliography}{56}
\expandafter\ifx\csname natexlab\endcsname\relax\def\natexlab#1{#1}\fi
\expandafter\ifx\csname bibnamefont\endcsname\relax
  \def\bibnamefont#1{#1}\fi
\expandafter\ifx\csname bibfnamefont\endcsname\relax
  \def\bibfnamefont#1{#1}\fi
\expandafter\ifx\csname citenamefont\endcsname\relax
  \def\citenamefont#1{#1}\fi
\expandafter\ifx\csname url\endcsname\relax
  \def\url#1{\texttt{#1}}\fi
\expandafter\ifx\csname urlprefix\endcsname\relax\def\urlprefix{URL }\fi
\providecommand{\bibinfo}[2]{#2}
\providecommand{\eprint}[2][]{\url{#2}}

\bibitem[{\citenamefont{Shekhter et~al.}(2013)\citenamefont{Shekhter, Ramshaw,
  Liang, Hardy, Bonn, Balakirev, McDonald, Betts, Riggs, and
  Migliori}}]{Shekhter:2013eh}
\bibinfo{author}{\bibfnamefont{A.}~\bibnamefont{Shekhter}},
  \bibinfo{author}{\bibfnamefont{B.~J.} \bibnamefont{Ramshaw}},
  \bibinfo{author}{\bibfnamefont{R.}~\bibnamefont{Liang}},
  \bibinfo{author}{\bibfnamefont{W.~N.} \bibnamefont{Hardy}},
  \bibinfo{author}{\bibfnamefont{D.~A.} \bibnamefont{Bonn}},
  \bibinfo{author}{\bibfnamefont{F.~F.} \bibnamefont{Balakirev}},
  \bibinfo{author}{\bibfnamefont{R.~D.} \bibnamefont{McDonald}},
  \bibinfo{author}{\bibfnamefont{J.~B.} \bibnamefont{Betts}},
  \bibinfo{author}{\bibfnamefont{S.~C.} \bibnamefont{Riggs}}, \bibnamefont{and}
  \bibinfo{author}{\bibfnamefont{A.}~\bibnamefont{Migliori}},
  \bibinfo{journal}{Nature (London)} \textbf{\bibinfo{volume}{498}},
  \bibinfo{pages}{75} (\bibinfo{year}{2013}).

\bibitem[{\citenamefont{Fauqu\'{e} et~al.}(2006)\citenamefont{Fauqu\'{e},
  Sidis, Hinkov, Pailh\`{e}s, Lin, Chaud, and Bourges}}]{Fauque2006}
\bibinfo{author}{\bibfnamefont{B.}~\bibnamefont{Fauqu\'{e}}},
  \bibinfo{author}{\bibfnamefont{Y.}~\bibnamefont{Sidis}},
  \bibinfo{author}{\bibfnamefont{V.}~\bibnamefont{Hinkov}},
  \bibinfo{author}{\bibfnamefont{S.}~\bibnamefont{Pailh\`{e}s}},
  \bibinfo{author}{\bibfnamefont{C.~T.} \bibnamefont{Lin}},
  \bibinfo{author}{\bibfnamefont{X.}~\bibnamefont{Chaud}}, \bibnamefont{and}
  \bibinfo{author}{\bibfnamefont{P.}~\bibnamefont{Bourges}},
  \bibinfo{journal}{Phys. Rev. Lett.} \textbf{\bibinfo{volume}{96}},
  \bibinfo{pages}{197001} (\bibinfo{year}{2006}).

\bibitem[{\citenamefont{Li et~al.}(2008)\citenamefont{Li, Baledent, Barisic,
  Cho, Fauque, Sidis, Yu, Zhao, Bourges, and Greven}}]{Li2008}
\bibinfo{author}{\bibfnamefont{Y.}~\bibnamefont{Li}},
  \bibinfo{author}{\bibfnamefont{V.}~\bibnamefont{Baledent}},
  \bibinfo{author}{\bibfnamefont{N.}~\bibnamefont{Barisic}},
  \bibinfo{author}{\bibfnamefont{Y.}~\bibnamefont{Cho}},
  \bibinfo{author}{\bibfnamefont{B.}~\bibnamefont{Fauque}},
  \bibinfo{author}{\bibfnamefont{Y.}~\bibnamefont{Sidis}},
  \bibinfo{author}{\bibfnamefont{G.}~\bibnamefont{Yu}},
  \bibinfo{author}{\bibfnamefont{X.}~\bibnamefont{Zhao}},
  \bibinfo{author}{\bibfnamefont{P.}~\bibnamefont{Bourges}}, \bibnamefont{and}
  \bibinfo{author}{\bibfnamefont{M.}~\bibnamefont{Greven}},
  \bibinfo{journal}{Nature} \textbf{\bibinfo{volume}{455}},
  \bibinfo{pages}{372} (\bibinfo{year}{2008}).

\bibitem[{\citenamefont{Li et~al.}(2011)\citenamefont{Li, Bal\'edent,
  Bari\ifmmode \check{s}\else \v{s}\fi{}i\ifmmode~\acute{c}\else \'{c}\fi{},
  Cho, Sidis, Yu, Zhao, Bourges, and Greven}}]{Li2011}
\bibinfo{author}{\bibfnamefont{Y.}~\bibnamefont{Li}},
  \bibinfo{author}{\bibfnamefont{V.}~\bibnamefont{Bal\'edent}},
  \bibinfo{author}{\bibfnamefont{N.}~\bibnamefont{Bari\ifmmode \check{s}\else
  \v{s}\fi{}i\ifmmode~\acute{c}\else \'{c}\fi{}}},
  \bibinfo{author}{\bibfnamefont{Y.~C.} \bibnamefont{Cho}},
  \bibinfo{author}{\bibfnamefont{Y.}~\bibnamefont{Sidis}},
  \bibinfo{author}{\bibfnamefont{G.}~\bibnamefont{Yu}},
  \bibinfo{author}{\bibfnamefont{X.}~\bibnamefont{Zhao}},
  \bibinfo{author}{\bibfnamefont{P.}~\bibnamefont{Bourges}}, \bibnamefont{and}
  \bibinfo{author}{\bibfnamefont{M.}~\bibnamefont{Greven}},
  \bibinfo{journal}{Phys. Rev. B} \textbf{\bibinfo{volume}{84}},
  \bibinfo{pages}{224508} (\bibinfo{year}{2011}).

\bibitem[{\citenamefont{Sidis and Bourges}(2013)}]{Sidis2013}
\bibinfo{author}{\bibfnamefont{Y.}~\bibnamefont{Sidis}} \bibnamefont{and}
  \bibinfo{author}{\bibfnamefont{P.}~\bibnamefont{Bourges}},
  \bibinfo{journal}{J. Phys.: Conf. Ser.} \textbf{\bibinfo{volume}{449}},
  \bibinfo{pages}{012012} (\bibinfo{year}{2013}).

\bibitem[{\citenamefont{Varma}(2006)}]{Varma2006}
\bibinfo{author}{\bibfnamefont{C.~M.} \bibnamefont{Varma}},
  \bibinfo{journal}{Phys. Rev. B} \textbf{\bibinfo{volume}{73}},
  \bibinfo{pages}{155113} (\bibinfo{year}{2006}).

\bibitem[{\citenamefont{Varma}(2014)}]{Varma:2014id}
\bibinfo{author}{\bibfnamefont{C.~M.} \bibnamefont{Varma}},
  \bibinfo{journal}{J. Phys.: Condens. Matter} \textbf{\bibinfo{volume}{26}},
  \bibinfo{pages}{505701} (\bibinfo{year}{2014}).

\bibitem[{\citenamefont{Weber et~al.}(2009)\citenamefont{Weber, L\"{a}uchli,
  Mila, and Giamarchi}}]{Weber2009}
\bibinfo{author}{\bibfnamefont{C.}~\bibnamefont{Weber}},
  \bibinfo{author}{\bibfnamefont{A.}~\bibnamefont{L\"{a}uchli}},
  \bibinfo{author}{\bibfnamefont{F.}~\bibnamefont{Mila}}, \bibnamefont{and}
  \bibinfo{author}{\bibfnamefont{T.}~\bibnamefont{Giamarchi}},
  \bibinfo{journal}{Phys. Rev. Lett.} \textbf{\bibinfo{volume}{102}},
  \bibinfo{pages}{017005} (\bibinfo{year}{2009}).

\bibitem[{\citenamefont{Weber et~al.}(2014)\citenamefont{Weber, Giamarchi, and
  Varma}}]{Weber2014}
\bibinfo{author}{\bibfnamefont{C.}~\bibnamefont{Weber}},
  \bibinfo{author}{\bibfnamefont{T.}~\bibnamefont{Giamarchi}},
  \bibnamefont{and} \bibinfo{author}{\bibfnamefont{C.~M.} \bibnamefont{Varma}},
  \bibinfo{journal}{Phys. Rev. Lett.} \textbf{\bibinfo{volume}{112}},
  \bibinfo{pages}{117001} (\bibinfo{year}{2014}).

\bibitem[{\citenamefont{Thomale and Greiter}(2008)}]{Thomale2008}
\bibinfo{author}{\bibfnamefont{R.}~\bibnamefont{Thomale}} \bibnamefont{and}
  \bibinfo{author}{\bibfnamefont{M.}~\bibnamefont{Greiter}},
  \bibinfo{journal}{Phys. Rev. B} \textbf{\bibinfo{volume}{77}},
  \bibinfo{pages}{094511} (\bibinfo{year}{2008}).

\bibitem[{\citenamefont{Kung et~al.}(2014)\citenamefont{Kung, Chen, Moritz,
  Johnston, Thomale, and Devereaux}}]{Kung2014}
\bibinfo{author}{\bibfnamefont{Y.~F.} \bibnamefont{Kung}},
  \bibinfo{author}{\bibfnamefont{C.-C.} \bibnamefont{Chen}},
  \bibinfo{author}{\bibfnamefont{B.}~\bibnamefont{Moritz}},
  \bibinfo{author}{\bibfnamefont{S.}~\bibnamefont{Johnston}},
  \bibinfo{author}{\bibfnamefont{R.}~\bibnamefont{Thomale}}, \bibnamefont{and}
  \bibinfo{author}{\bibfnamefont{T.~P.} \bibnamefont{Devereaux}},
  \bibinfo{journal}{Phys. Rev. B} \textbf{\bibinfo{volume}{90}},
  \bibinfo{pages}{224507} (\bibinfo{year}{2014}).

\bibitem[{\citenamefont{Xia et~al.}(2008)\citenamefont{Xia, Schemm, Deutscher,
  Kivelson, Bonn, Hardy, Liang, Siemons, Koster, Fejer, and
  Kapitulnik}}]{Xia2008}
\bibinfo{author}{\bibfnamefont{J.}~\bibnamefont{Xia}},
  \bibinfo{author}{\bibfnamefont{E.}~\bibnamefont{Schemm}},
  \bibinfo{author}{\bibfnamefont{G.}~\bibnamefont{Deutscher}},
  \bibinfo{author}{\bibfnamefont{S.~A.} \bibnamefont{Kivelson}},
  \bibinfo{author}{\bibfnamefont{D.~A.} \bibnamefont{Bonn}},
  \bibinfo{author}{\bibfnamefont{W.~N.} \bibnamefont{Hardy}},
  \bibinfo{author}{\bibfnamefont{R.}~\bibnamefont{Liang}},
  \bibinfo{author}{\bibfnamefont{W.}~\bibnamefont{Siemons}},
  \bibinfo{author}{\bibfnamefont{G.}~\bibnamefont{Koster}},
  \bibinfo{author}{\bibfnamefont{M.~M.} \bibnamefont{Fejer}}, \bibnamefont{and}
  \bibinfo{author}{\bibfnamefont{A.}~\bibnamefont{Kapitulnik}},
  \bibinfo{journal}{Phys. Rev. Lett.} \textbf{\bibinfo{volume}{100}},
  \bibinfo{pages}{127002} (\bibinfo{year}{2008}).

\bibitem[{\citenamefont{He et~al.}(2011)\citenamefont{He, Hashimoto,
  Karapetyan, Koralek, Hinton, Testaud, Nathan, Yoshida, Yao, Tanaka,
  Meevasana, Moore, Lu, Mo, Ishikado, Eisaki, Hussain, Devereaux, Kivelson,
  Orenstein, Kapitulnik, and Shen}}]{RHe2011}
\bibinfo{author}{\bibfnamefont{R.-H.} \bibnamefont{He}},
  \bibinfo{author}{\bibfnamefont{M.}~\bibnamefont{Hashimoto}},
  \bibinfo{author}{\bibfnamefont{H.}~\bibnamefont{Karapetyan}},
  \bibinfo{author}{\bibfnamefont{J.~D.} \bibnamefont{Koralek}},
  \bibinfo{author}{\bibfnamefont{J.~P.} \bibnamefont{Hinton}},
  \bibinfo{author}{\bibfnamefont{J.~P.} \bibnamefont{Testaud}},
  \bibinfo{author}{\bibfnamefont{V.}~\bibnamefont{Nathan}},
  \bibinfo{author}{\bibfnamefont{Y.}~\bibnamefont{Yoshida}},
  \bibinfo{author}{\bibfnamefont{H.}~\bibnamefont{Yao}},
  \bibinfo{author}{\bibfnamefont{K.}~\bibnamefont{Tanaka}},
  \bibinfo{author}{\bibfnamefont{W.}~\bibnamefont{Meevasana}},
  \bibinfo{author}{\bibfnamefont{R.~G.} \bibnamefont{Moore}},
  \bibinfo{author}{\bibfnamefont{D.~H.} \bibnamefont{Lu}},
  \bibinfo{author}{\bibfnamefont{S.-K.} \bibnamefont{Mo}},
  \bibinfo{author}{\bibfnamefont{M.}~\bibnamefont{Ishikado}},
  \bibinfo{author}{\bibfnamefont{H.}~\bibnamefont{Eisaki}},
  \bibinfo{author}{\bibfnamefont{Z.}~\bibnamefont{Hussain}},
  \bibinfo{author}{\bibfnamefont{T.~P.} \bibnamefont{Devereaux}},
  \bibinfo{author}{\bibfnamefont{S.~A.} \bibnamefont{Kivelson}},
  \bibinfo{author}{\bibfnamefont{J.}~\bibnamefont{Orenstein}},
  \bibinfo{author}{\bibfnamefont{A.}~\bibnamefont{Kapitulnik}},
  \bibnamefont{and} \bibinfo{author}{\bibfnamefont{Z.-X.} \bibnamefont{Shen}},
  \bibinfo{journal}{Science} \textbf{\bibinfo{volume}{331}},
  \bibinfo{pages}{1579} (\bibinfo{year}{2011}).

\bibitem[{\citenamefont{Wang et~al.}(2014)\citenamefont{Wang, Chubukov, and
  Nandkishore}}]{Wang2014}
\bibinfo{author}{\bibfnamefont{Y.}~\bibnamefont{Wang}},
  \bibinfo{author}{\bibfnamefont{A.}~\bibnamefont{Chubukov}}, \bibnamefont{and}
  \bibinfo{author}{\bibfnamefont{R.}~\bibnamefont{Nandkishore}},
  \bibinfo{journal}{Phys. Rev. B} \textbf{\bibinfo{volume}{90}},
  \bibinfo{pages}{205130} (\bibinfo{year}{2014}).

\bibitem[{\citenamefont{Wu et~al.}(2011)\citenamefont{Wu, Mayaffre, Kr{\"a}mer,
  Horvati{\'c}, Berthier, Hardy, Liang, Bonn, and Julien}}]{Wu2011}
\bibinfo{author}{\bibfnamefont{T.}~\bibnamefont{Wu}},
  \bibinfo{author}{\bibfnamefont{H.}~\bibnamefont{Mayaffre}},
  \bibinfo{author}{\bibfnamefont{S.}~\bibnamefont{Kr{\"a}mer}},
  \bibinfo{author}{\bibfnamefont{M.}~\bibnamefont{Horvati{\'c}}},
  \bibinfo{author}{\bibfnamefont{C.}~\bibnamefont{Berthier}},
  \bibinfo{author}{\bibfnamefont{W.~N.} \bibnamefont{Hardy}},
  \bibinfo{author}{\bibfnamefont{R.}~\bibnamefont{Liang}},
  \bibinfo{author}{\bibfnamefont{D.~A.} \bibnamefont{Bonn}}, \bibnamefont{and}
  \bibinfo{author}{\bibfnamefont{M.-H.} \bibnamefont{Julien}},
  \bibinfo{journal}{Nature (London)} \textbf{\bibinfo{volume}{477}},
  \bibinfo{pages}{191} (\bibinfo{year}{2011}).

\bibitem[{\citenamefont{Ghiringhelli et~al.}(2012)\citenamefont{Ghiringhelli,
  Le~Tacon, Minola, Blanco-Canosa, Mazzoli, Brookes, De~Luca, Frano, Hawthorn,
  He, Loew, Sala, Peets, Salluzzo, Schierle, Sutarto, Sawatzky, Weschke,
  Keimer, and Braicovich}}]{Ghiringelli2012}
\bibinfo{author}{\bibfnamefont{G.}~\bibnamefont{Ghiringhelli}},
  \bibinfo{author}{\bibfnamefont{M.}~\bibnamefont{Le~Tacon}},
  \bibinfo{author}{\bibfnamefont{M.}~\bibnamefont{Minola}},
  \bibinfo{author}{\bibfnamefont{S.}~\bibnamefont{Blanco-Canosa}},
  \bibinfo{author}{\bibfnamefont{C.}~\bibnamefont{Mazzoli}},
  \bibinfo{author}{\bibfnamefont{N.~B.} \bibnamefont{Brookes}},
  \bibinfo{author}{\bibfnamefont{G.~M.} \bibnamefont{De~Luca}},
  \bibinfo{author}{\bibfnamefont{A.}~\bibnamefont{Frano}},
  \bibinfo{author}{\bibfnamefont{D.~G.} \bibnamefont{Hawthorn}},
  \bibinfo{author}{\bibfnamefont{F.}~\bibnamefont{He}},
  \bibinfo{author}{\bibfnamefont{T.}~\bibnamefont{Loew}},
  \bibinfo{author}{\bibfnamefont{M.~M.} \bibnamefont{Sala}},
  \bibinfo{author}{\bibfnamefont{D.~C.} \bibnamefont{Peets}},
  \bibinfo{author}{\bibfnamefont{M.}~\bibnamefont{Salluzzo}},
  \bibinfo{author}{\bibfnamefont{E.}~\bibnamefont{Schierle}},
  \bibinfo{author}{\bibfnamefont{R.}~\bibnamefont{Sutarto}},
  \bibinfo{author}{\bibfnamefont{G.~A.} \bibnamefont{Sawatzky}},
  \bibinfo{author}{\bibfnamefont{E.}~\bibnamefont{Weschke}},
  \bibinfo{author}{\bibfnamefont{B.}~\bibnamefont{Keimer}}, \bibnamefont{and}
  \bibinfo{author}{\bibfnamefont{L.}~\bibnamefont{Braicovich}},
  \bibinfo{journal}{Science} \textbf{\bibinfo{volume}{337}},
  \bibinfo{pages}{821} (\bibinfo{year}{2012}).

\bibitem[{\citenamefont{Chang et~al.}(2012)\citenamefont{Chang, Blackburn,
  Holmes, Christensen, Larsen, Mesot, Liang, Bonn, Hardy, Watenphul,
  Zimmermann, Forgan, and Hayden}}]{Chang2012vf}
\bibinfo{author}{\bibfnamefont{J.}~\bibnamefont{Chang}},
  \bibinfo{author}{\bibfnamefont{E.}~\bibnamefont{Blackburn}},
  \bibinfo{author}{\bibfnamefont{A.~T.} \bibnamefont{Holmes}},
  \bibinfo{author}{\bibfnamefont{N.~B.} \bibnamefont{Christensen}},
  \bibinfo{author}{\bibfnamefont{J.}~\bibnamefont{Larsen}},
  \bibinfo{author}{\bibfnamefont{J.}~\bibnamefont{Mesot}},
  \bibinfo{author}{\bibfnamefont{R.}~\bibnamefont{Liang}},
  \bibinfo{author}{\bibfnamefont{D.~A.} \bibnamefont{Bonn}},
  \bibinfo{author}{\bibfnamefont{W.~N.} \bibnamefont{Hardy}},
  \bibinfo{author}{\bibfnamefont{A.}~\bibnamefont{Watenphul}},
  \bibinfo{author}{\bibfnamefont{M.~v.} \bibnamefont{Zimmermann}},
  \bibinfo{author}{\bibfnamefont{E.~M.} \bibnamefont{Forgan}},
  \bibnamefont{and} \bibinfo{author}{\bibfnamefont{S.~M.}
  \bibnamefont{Hayden}}, \bibinfo{journal}{Nature Phys.}
  \textbf{\bibinfo{volume}{8}}, \bibinfo{pages}{871} (\bibinfo{year}{2012}).

\bibitem[{\citenamefont{da~Silva~Neto et~al.}(2014)\citenamefont{da~Silva~Neto,
  Aynajian, Frano, Comin, Schierle, Weschke, Gyenis, Wen, Schneeloch, Xu, Ono,
  Gu, Le~Tacon, and Yazdani}}]{Neto2014}
\bibinfo{author}{\bibfnamefont{E.~H.} \bibnamefont{da~Silva~Neto}},
  \bibinfo{author}{\bibfnamefont{P.}~\bibnamefont{Aynajian}},
  \bibinfo{author}{\bibfnamefont{A.}~\bibnamefont{Frano}},
  \bibinfo{author}{\bibfnamefont{R.}~\bibnamefont{Comin}},
  \bibinfo{author}{\bibfnamefont{E.}~\bibnamefont{Schierle}},
  \bibinfo{author}{\bibfnamefont{E.}~\bibnamefont{Weschke}},
  \bibinfo{author}{\bibfnamefont{A.}~\bibnamefont{Gyenis}},
  \bibinfo{author}{\bibfnamefont{J.}~\bibnamefont{Wen}},
  \bibinfo{author}{\bibfnamefont{J.}~\bibnamefont{Schneeloch}},
  \bibinfo{author}{\bibfnamefont{Z.}~\bibnamefont{Xu}},
  \bibinfo{author}{\bibfnamefont{S.}~\bibnamefont{Ono}},
  \bibinfo{author}{\bibfnamefont{G.}~\bibnamefont{Gu}},
  \bibinfo{author}{\bibfnamefont{M.}~\bibnamefont{Le~Tacon}}, \bibnamefont{and}
  \bibinfo{author}{\bibfnamefont{A.}~\bibnamefont{Yazdani}},
  \bibinfo{journal}{Science} \textbf{\bibinfo{volume}{343}},
  \bibinfo{pages}{393} (\bibinfo{year}{2014}).

\bibitem[{\citenamefont{Comin et~al.}(2014)\citenamefont{Comin, Frano, Yee,
  Yoshida, Eisaki, Schierle, Weschke, Sutarto, He, Soumyanarayanan, He,
  Le~Tacon, Elfimov, Hoffman, Sawatzky, Keimer, and Damascelli}}]{Comin2014}
\bibinfo{author}{\bibfnamefont{R.}~\bibnamefont{Comin}},
  \bibinfo{author}{\bibfnamefont{A.}~\bibnamefont{Frano}},
  \bibinfo{author}{\bibfnamefont{M.~M.} \bibnamefont{Yee}},
  \bibinfo{author}{\bibfnamefont{Y.}~\bibnamefont{Yoshida}},
  \bibinfo{author}{\bibfnamefont{H.}~\bibnamefont{Eisaki}},
  \bibinfo{author}{\bibfnamefont{E.}~\bibnamefont{Schierle}},
  \bibinfo{author}{\bibfnamefont{E.}~\bibnamefont{Weschke}},
  \bibinfo{author}{\bibfnamefont{R.}~\bibnamefont{Sutarto}},
  \bibinfo{author}{\bibfnamefont{F.}~\bibnamefont{He}},
  \bibinfo{author}{\bibfnamefont{A.}~\bibnamefont{Soumyanarayanan}},
  \bibinfo{author}{\bibfnamefont{Y.}~\bibnamefont{He}},
  \bibinfo{author}{\bibfnamefont{M.}~\bibnamefont{Le~Tacon}},
  \bibinfo{author}{\bibfnamefont{I.~S.} \bibnamefont{Elfimov}},
  \bibinfo{author}{\bibfnamefont{J.~E.} \bibnamefont{Hoffman}},
  \bibinfo{author}{\bibfnamefont{G.~A.} \bibnamefont{Sawatzky}},
  \bibinfo{author}{\bibfnamefont{B.}~\bibnamefont{Keimer}}, \bibnamefont{and}
  \bibinfo{author}{\bibfnamefont{A.}~\bibnamefont{Damascelli}},
  \bibinfo{journal}{Science} \textbf{\bibinfo{volume}{343}},
  \bibinfo{pages}{390} (\bibinfo{year}{2014}).

\bibitem[{\citenamefont{Kohsaka et~al.}(2007)\citenamefont{Kohsaka, Taylor,
  Fujita, Schmidt, Lupien, Hanaguri, Azuma, Takano, Eisaki, Takagi, Uchida, and
  Davis}}]{Kohsaka2007}
\bibinfo{author}{\bibfnamefont{Y.}~\bibnamefont{Kohsaka}},
  \bibinfo{author}{\bibfnamefont{C.}~\bibnamefont{Taylor}},
  \bibinfo{author}{\bibfnamefont{K.}~\bibnamefont{Fujita}},
  \bibinfo{author}{\bibfnamefont{A.}~\bibnamefont{Schmidt}},
  \bibinfo{author}{\bibfnamefont{C.}~\bibnamefont{Lupien}},
  \bibinfo{author}{\bibfnamefont{T.}~\bibnamefont{Hanaguri}},
  \bibinfo{author}{\bibfnamefont{M.}~\bibnamefont{Azuma}},
  \bibinfo{author}{\bibfnamefont{M.}~\bibnamefont{Takano}},
  \bibinfo{author}{\bibfnamefont{H.}~\bibnamefont{Eisaki}},
  \bibinfo{author}{\bibfnamefont{H.}~\bibnamefont{Takagi}},
  \bibinfo{author}{\bibfnamefont{S.}~\bibnamefont{Uchida}}, \bibnamefont{and}
  \bibinfo{author}{\bibfnamefont{J.~C.} \bibnamefont{Davis}},
  \bibinfo{journal}{Science} \textbf{\bibinfo{volume}{315}},
  \bibinfo{pages}{1380} (\bibinfo{year}{2007}).

\bibitem[{\citenamefont{Wise et~al.}(2008)\citenamefont{Wise, Boyer,
  Chatterjee, Kondo, Takeuchi, Ikuta, Wang, and Hudson}}]{Wise2008}
\bibinfo{author}{\bibfnamefont{W.~D.} \bibnamefont{Wise}},
  \bibinfo{author}{\bibfnamefont{M.~C.} \bibnamefont{Boyer}},
  \bibinfo{author}{\bibfnamefont{K.}~\bibnamefont{Chatterjee}},
  \bibinfo{author}{\bibfnamefont{T.}~\bibnamefont{Kondo}},
  \bibinfo{author}{\bibfnamefont{T.}~\bibnamefont{Takeuchi}},
  \bibinfo{author}{\bibfnamefont{H.}~\bibnamefont{Ikuta}},
  \bibinfo{author}{\bibfnamefont{Y.}~\bibnamefont{Wang}}, \bibnamefont{and}
  \bibinfo{author}{\bibfnamefont{E.~W.} \bibnamefont{Hudson}},
  \bibinfo{journal}{Nat. Phys.} \textbf{\bibinfo{volume}{4}},
  \bibinfo{pages}{696} (\bibinfo{year}{2008}).

\bibitem[{\citenamefont{Lawler et~al.}(2010)\citenamefont{Lawler, Fujita, Lee,
  Schmidt, Kohsaka, Kim, Eisaki, Uchida, Davis, Sethna, and Kim}}]{Lawler2010}
\bibinfo{author}{\bibfnamefont{M.~J.} \bibnamefont{Lawler}},
  \bibinfo{author}{\bibfnamefont{K.}~\bibnamefont{Fujita}},
  \bibinfo{author}{\bibfnamefont{J.}~\bibnamefont{Lee}},
  \bibinfo{author}{\bibfnamefont{A.~R.} \bibnamefont{Schmidt}},
  \bibinfo{author}{\bibfnamefont{Y.}~\bibnamefont{Kohsaka}},
  \bibinfo{author}{\bibfnamefont{C.~K.} \bibnamefont{Kim}},
  \bibinfo{author}{\bibfnamefont{H.}~\bibnamefont{Eisaki}},
  \bibinfo{author}{\bibfnamefont{S.}~\bibnamefont{Uchida}},
  \bibinfo{author}{\bibfnamefont{J.~C.} \bibnamefont{Davis}},
  \bibinfo{author}{\bibfnamefont{J.~P.} \bibnamefont{Sethna}},
  \bibnamefont{and} \bibinfo{author}{\bibfnamefont{E.-A.} \bibnamefont{Kim}},
  \bibinfo{journal}{Nature (London)} \textbf{\bibinfo{volume}{466}},
  \bibinfo{pages}{347} (\bibinfo{year}{2010}).

\bibitem[{\citenamefont{{Wu} et~al.}(2015)\citenamefont{{Wu}, {Mayaffre},
  {Kr{\"a}mer}, {Horvati{\'c}}, {Berthier}, {Hardy}, {Liang}, {Bonn}, and
  {Julien}}}]{Wu2015}
\bibinfo{author}{\bibfnamefont{T.}~\bibnamefont{{Wu}}},
  \bibinfo{author}{\bibfnamefont{H.}~\bibnamefont{{Mayaffre}}},
  \bibinfo{author}{\bibfnamefont{S.}~\bibnamefont{{Kr{\"a}mer}}},
  \bibinfo{author}{\bibfnamefont{M.}~\bibnamefont{{Horvati{\'c}}}},
  \bibinfo{author}{\bibfnamefont{C.}~\bibnamefont{{Berthier}}},
  \bibinfo{author}{\bibfnamefont{W.~N.} \bibnamefont{{Hardy}}},
  \bibinfo{author}{\bibfnamefont{R.}~\bibnamefont{{Liang}}},
  \bibinfo{author}{\bibfnamefont{D.~A.} \bibnamefont{{Bonn}}},
  \bibnamefont{and} \bibinfo{author}{\bibfnamefont{M.-H.}
  \bibnamefont{{Julien}}}, \bibinfo{journal}{Nat. Commun.}
  \textbf{\bibinfo{volume}{6}} (\bibinfo{year}{2015}).

\bibitem[{\citenamefont{Wang and Chubukov}(2014)}]{Wang:2014fr}
\bibinfo{author}{\bibfnamefont{Y.}~\bibnamefont{Wang}} \bibnamefont{and}
  \bibinfo{author}{\bibfnamefont{A.}~\bibnamefont{Chubukov}},
  \bibinfo{journal}{Phys. Rev. B} \textbf{\bibinfo{volume}{90}},
  \bibinfo{pages}{035149} (\bibinfo{year}{2014}).

\bibitem[{\citenamefont{Lee}(2014)}]{Lee:2014ka}
\bibinfo{author}{\bibfnamefont{P.~A.} \bibnamefont{Lee}},
  \bibinfo{journal}{Phys. Rev. X} \textbf{\bibinfo{volume}{4}},
  \bibinfo{pages}{031017} (\bibinfo{year}{2014}).

\bibitem[{\citenamefont{Agterberg et~al.}(2015)\citenamefont{Agterberg,
  Melchert, and Kashyap}}]{Agterberg:2015bo}
\bibinfo{author}{\bibfnamefont{D.~F.} \bibnamefont{Agterberg}},
  \bibinfo{author}{\bibfnamefont{D.~S.} \bibnamefont{Melchert}},
  \bibnamefont{and} \bibinfo{author}{\bibfnamefont{M.~K.}
  \bibnamefont{Kashyap}}, \bibinfo{journal}{Phys. Rev. B}
  \textbf{\bibinfo{volume}{91}}, \bibinfo{pages}{054502}
  (\bibinfo{year}{2015}).

\bibitem[{\citenamefont{Efetov et~al.}(2013)\citenamefont{Efetov, Meier, and
  P\'{e}pin}}]{Efetov2013}
\bibinfo{author}{\bibfnamefont{K.~B.} \bibnamefont{Efetov}},
  \bibinfo{author}{\bibfnamefont{H.}~\bibnamefont{Meier}}, \bibnamefont{and}
  \bibinfo{author}{\bibfnamefont{C.}~\bibnamefont{P\'{e}pin}},
  \bibinfo{journal}{Nat. Phys.} \textbf{\bibinfo{volume}{9}},
  \bibinfo{pages}{442} (\bibinfo{year}{2013}).

\bibitem[{\citenamefont{P\'epin et~al.}(2014)\citenamefont{P\'epin,
  de~Carvalho, Kloss, and Montiel}}]{Pepin:2014tb}
\bibinfo{author}{\bibfnamefont{C.}~\bibnamefont{P\'epin}},
  \bibinfo{author}{\bibfnamefont{V.~S.} \bibnamefont{de~Carvalho}},
  \bibinfo{author}{\bibfnamefont{T.}~\bibnamefont{Kloss}}, \bibnamefont{and}
  \bibinfo{author}{\bibfnamefont{X.}~\bibnamefont{Montiel}},
  \bibinfo{journal}{Phys. Rev. B} \textbf{\bibinfo{volume}{90}},
  \bibinfo{pages}{195207} (\bibinfo{year}{2014}).

\bibitem[{\citenamefont{Chowdhury and Sachdev}(2014)}]{Chowdhury2014}
\bibinfo{author}{\bibfnamefont{D.}~\bibnamefont{Chowdhury}} \bibnamefont{and}
  \bibinfo{author}{\bibfnamefont{S.}~\bibnamefont{Sachdev}},
  \bibinfo{journal}{Phys. Rev. B} \textbf{\bibinfo{volume}{90}},
  \bibinfo{pages}{134516} (\bibinfo{year}{2014}).

\bibitem[{\citenamefont{Hayward et~al.}(2014)\citenamefont{Hayward, Hawthorn,
  Melko, and Sachdev}}]{Hayward:2014eo}
\bibinfo{author}{\bibfnamefont{L.~E.} \bibnamefont{Hayward}},
  \bibinfo{author}{\bibfnamefont{D.~G.} \bibnamefont{Hawthorn}},
  \bibinfo{author}{\bibfnamefont{R.~G.} \bibnamefont{Melko}}, \bibnamefont{and}
  \bibinfo{author}{\bibfnamefont{S.}~\bibnamefont{Sachdev}},
  \bibinfo{journal}{Science} \textbf{\bibinfo{volume}{343}},
  \bibinfo{pages}{1336} (\bibinfo{year}{2014}).

\bibitem[{\citenamefont{Achkar et~al.}(unpublished)\citenamefont{Achkar, He,
  Sutarto, McMahon, Zwiebler, Hucker, Gu, Liang, Bonn, Hardy, Geck, and
  Hawthorn}}]{Achkar2015}
\bibinfo{author}{\bibfnamefont{A.~J.} \bibnamefont{Achkar}},
  \bibinfo{author}{\bibfnamefont{F.}~\bibnamefont{He}},
  \bibinfo{author}{\bibfnamefont{R.}~\bibnamefont{Sutarto}},
  \bibinfo{author}{\bibfnamefont{C.}~\bibnamefont{McMahon}},
  \bibinfo{author}{\bibfnamefont{M.}~\bibnamefont{Zwiebler}},
  \bibinfo{author}{\bibfnamefont{M.}~\bibnamefont{Hucker}},
  \bibinfo{author}{\bibfnamefont{G.~D.} \bibnamefont{Gu}},
  \bibinfo{author}{\bibfnamefont{R.}~\bibnamefont{Liang}},
  \bibinfo{author}{\bibfnamefont{D.~A.} \bibnamefont{Bonn}},
  \bibinfo{author}{\bibfnamefont{W.~N.} \bibnamefont{Hardy}},
  \bibinfo{author}{\bibfnamefont{J.}~\bibnamefont{Geck}}, \bibnamefont{and}
  \bibinfo{author}{\bibfnamefont{D.~G.} \bibnamefont{Hawthorn}},
  \bibinfo{journal}{arXiv:1409.6787}  (\bibinfo{year}{unpublished}).

\bibitem[{\citenamefont{{Comin} et~al.}(2015)\citenamefont{{Comin}, {Sutarto},
  {He}, {da Silva Neto}, {Chauviere}, {Fra{\~n}o}, {Liang}, {Hardy}, {Bonn},
  {Yoshida}, {Eisaki}, {Achkar}, {Hawthorn}, {Keimer}, {Sawatzky}, and
  {Damascelli}}}]{Comin2015}
\bibinfo{author}{\bibfnamefont{R.}~\bibnamefont{{Comin}}},
  \bibinfo{author}{\bibfnamefont{R.}~\bibnamefont{{Sutarto}}},
  \bibinfo{author}{\bibfnamefont{F.}~\bibnamefont{{He}}},
  \bibinfo{author}{\bibfnamefont{E.~H.} \bibnamefont{{da Silva Neto}}},
  \bibinfo{author}{\bibfnamefont{L.}~\bibnamefont{{Chauviere}}},
  \bibinfo{author}{\bibfnamefont{A.}~\bibnamefont{{Fra{\~n}o}}},
  \bibinfo{author}{\bibfnamefont{R.}~\bibnamefont{{Liang}}},
  \bibinfo{author}{\bibfnamefont{W.~N.} \bibnamefont{{Hardy}}},
  \bibinfo{author}{\bibfnamefont{D.~A.} \bibnamefont{{Bonn}}},
  \bibinfo{author}{\bibfnamefont{Y.}~\bibnamefont{{Yoshida}}},
  \bibinfo{author}{\bibfnamefont{H.}~\bibnamefont{{Eisaki}}},
  \bibinfo{author}{\bibfnamefont{A.~J.} \bibnamefont{{Achkar}}},
  \bibinfo{author}{\bibfnamefont{D.~G.} \bibnamefont{{Hawthorn}}},
  \bibinfo{author}{\bibfnamefont{B.}~\bibnamefont{{Keimer}}},
  \bibinfo{author}{\bibfnamefont{G.~A.} \bibnamefont{{Sawatzky}}},
  \bibnamefont{and}
  \bibinfo{author}{\bibfnamefont{A.}~\bibnamefont{{Damascelli}}},
  \bibinfo{journal}{Nat. Mater.} \textbf{\bibinfo{volume}{14}},
  \bibinfo{pages}{796} (\bibinfo{year}{2015}).

\bibitem[{\citenamefont{Fischer and Kim}(2011)}]{Fischer2011}
\bibinfo{author}{\bibfnamefont{M.~H.} \bibnamefont{Fischer}} \bibnamefont{and}
  \bibinfo{author}{\bibfnamefont{E.-A.} \bibnamefont{Kim}},
  \bibinfo{journal}{Phys. Rev. B} \textbf{\bibinfo{volume}{84}},
  \bibinfo{pages}{144502} (\bibinfo{year}{2011}).

\bibitem[{\citenamefont{Bulut et~al.}(2013)\citenamefont{Bulut, Atkinson, and
  Kampf}}]{Bulut2013}
\bibinfo{author}{\bibfnamefont{S.}~\bibnamefont{Bulut}},
  \bibinfo{author}{\bibfnamefont{W.~A.} \bibnamefont{Atkinson}},
  \bibnamefont{and} \bibinfo{author}{\bibfnamefont{A.~P.} \bibnamefont{Kampf}},
  \bibinfo{journal}{Phys. Rev. B} \textbf{\bibinfo{volume}{88}}
  (\bibinfo{year}{2013}).

\bibitem[{\citenamefont{Atkinson et~al.}(2015)\citenamefont{Atkinson, Kampf,
  and Bulut}}]{Atkinson2014}
\bibinfo{author}{\bibfnamefont{W.~A.} \bibnamefont{Atkinson}},
  \bibinfo{author}{\bibfnamefont{A.~P.} \bibnamefont{Kampf}}, \bibnamefont{and}
  \bibinfo{author}{\bibfnamefont{S.}~\bibnamefont{Bulut}},
  \bibinfo{journal}{New J. Phys.} \textbf{\bibinfo{volume}{17}},
  \bibinfo{pages}{013025} (\bibinfo{year}{2015}).

\bibitem[{\citenamefont{Chakravarty
  et~al.}(2001{\natexlab{a}})\citenamefont{Chakravarty, Laughlin, Morr, and
  Nayak}}]{Chakravarty2001}
\bibinfo{author}{\bibfnamefont{S.}~\bibnamefont{Chakravarty}},
  \bibinfo{author}{\bibfnamefont{R.~B.} \bibnamefont{Laughlin}},
  \bibinfo{author}{\bibfnamefont{D.~K.} \bibnamefont{Morr}}, \bibnamefont{and}
  \bibinfo{author}{\bibfnamefont{C.}~\bibnamefont{Nayak}},
  \bibinfo{journal}{Phys. Rev. B} \textbf{\bibinfo{volume}{63}},
  \bibinfo{pages}{094503} (\bibinfo{year}{2001}{\natexlab{a}}).

\bibitem[{\citenamefont{Chakravarty et~al.}(2004)\citenamefont{Chakravarty,
  Kee, and V\"olker}}]{Chakravarty2004}
\bibinfo{author}{\bibfnamefont{S.}~\bibnamefont{Chakravarty}},
  \bibinfo{author}{\bibfnamefont{H.-Y.} \bibnamefont{Kee}}, \bibnamefont{and}
  \bibinfo{author}{\bibfnamefont{K.}~\bibnamefont{V\"olker}},
  \bibinfo{journal}{Nature (London)} \textbf{\bibinfo{volume}{428}},
  \bibinfo{pages}{53} (\bibinfo{year}{2004}).

\bibitem[{\citenamefont{Emery}(1987)}]{Emery1987}
\bibinfo{author}{\bibfnamefont{V.~J.} \bibnamefont{Emery}},
  \bibinfo{journal}{Phys. Rev. Lett.} \textbf{\bibinfo{volume}{58}},
  \bibinfo{pages}{2794} (\bibinfo{year}{1987}).

\bibitem[{\citenamefont{Hybertsen et~al.}(1989)\citenamefont{Hybertsen,
  Schl\"uter, and Christensen}}]{Hybertsen1989}
\bibinfo{author}{\bibfnamefont{M.~S.} \bibnamefont{Hybertsen}},
  \bibinfo{author}{\bibfnamefont{M.}~\bibnamefont{Schl\"uter}},
  \bibnamefont{and} \bibinfo{author}{\bibfnamefont{N.~E.}
  \bibnamefont{Christensen}}, \bibinfo{journal}{Phys. Rev. B}
  \textbf{\bibinfo{volume}{39}}, \bibinfo{pages}{9028} (\bibinfo{year}{1989}).

\bibitem[{\citenamefont{Littlewood et~al.}(1989)\citenamefont{Littlewood,
  Varma, Schmitt-Rink, and Abrahams}}]{Littlewood1989}
\bibinfo{author}{\bibfnamefont{P.~B.} \bibnamefont{Littlewood}},
  \bibinfo{author}{\bibfnamefont{C.~M.} \bibnamefont{Varma}},
  \bibinfo{author}{\bibfnamefont{S.}~\bibnamefont{Schmitt-Rink}},
  \bibnamefont{and} \bibinfo{author}{\bibfnamefont{E.}~\bibnamefont{Abrahams}},
  \bibinfo{journal}{Phys. Rev. B} \textbf{\bibinfo{volume}{39}},
  \bibinfo{pages}{12371} (\bibinfo{year}{1989}).

\bibitem[{\citenamefont{Hsu et~al.}(1991)\citenamefont{Hsu, Marston, and
  Affleck}}]{Hsu1991}
\bibinfo{author}{\bibfnamefont{T.~C.} \bibnamefont{Hsu}},
  \bibinfo{author}{\bibfnamefont{J.~B.} \bibnamefont{Marston}},
  \bibnamefont{and} \bibinfo{author}{\bibfnamefont{I.}~\bibnamefont{Affleck}},
  \bibinfo{journal}{Phys. Rev. B} \textbf{\bibinfo{volume}{43}},
  \bibinfo{pages}{2866} (\bibinfo{year}{1991}).

\bibitem[{\citenamefont{Ivanov et~al.}(2000)\citenamefont{Ivanov, Lee, and
  Wen}}]{Ivanov2000}
\bibinfo{author}{\bibfnamefont{D.~A.} \bibnamefont{Ivanov}},
  \bibinfo{author}{\bibfnamefont{P.~A.} \bibnamefont{Lee}}, \bibnamefont{and}
  \bibinfo{author}{\bibfnamefont{X.-G.} \bibnamefont{Wen}},
  \bibinfo{journal}{Phys. Rev. Lett.} \textbf{\bibinfo{volume}{84}},
  \bibinfo{pages}{3958} (\bibinfo{year}{2000}).

\bibitem[{\citenamefont{Tsutsui et~al.}(2001)\citenamefont{Tsutsui, Poilblanc,
  and Capponi}}]{Tsutsui2002}
\bibinfo{author}{\bibfnamefont{K.}~\bibnamefont{Tsutsui}},
  \bibinfo{author}{\bibfnamefont{D.}~\bibnamefont{Poilblanc}},
  \bibnamefont{and} \bibinfo{author}{\bibfnamefont{S.}~\bibnamefont{Capponi}},
  \bibinfo{journal}{Phys. Rev. B} \textbf{\bibinfo{volume}{65}},
  \bibinfo{pages}{020406} (\bibinfo{year}{2001}).

\bibitem[{\citenamefont{He et~al.}(2012)\citenamefont{He, Moore, and
  Varma}}]{He2012b}
\bibinfo{author}{\bibfnamefont{Y.}~\bibnamefont{He}},
  \bibinfo{author}{\bibfnamefont{J.}~\bibnamefont{Moore}}, \bibnamefont{and}
  \bibinfo{author}{\bibfnamefont{C.~M.} \bibnamefont{Varma}},
  \bibinfo{journal}{Phys. Rev. B} \textbf{\bibinfo{volume}{85}},
  \bibinfo{pages}{155106} (\bibinfo{year}{2012}).

\bibitem[{\citenamefont{Pasanai and Atkinson}(2010)}]{Pasanai2010}
\bibinfo{author}{\bibfnamefont{K.}~\bibnamefont{Pasanai}} \bibnamefont{and}
  \bibinfo{author}{\bibfnamefont{W.~A.} \bibnamefont{Atkinson}},
  \bibinfo{journal}{Phys. Rev. B} \textbf{\bibinfo{volume}{81}},
  \bibinfo{pages}{134501} (\bibinfo{year}{2010}).

\bibitem[{\citenamefont{Laughlin}(2014)}]{Laughlin2014prb}
\bibinfo{author}{\bibfnamefont{R.~B.} \bibnamefont{Laughlin}},
  \bibinfo{journal}{Phys. Rev. B} \textbf{\bibinfo{volume}{89}},
  \bibinfo{pages}{035134} (\bibinfo{year}{2014}).

\bibitem[{\citenamefont{Dimov et~al.}(2008)\citenamefont{Dimov, Goswami, Jia,
  and Chakravarty}}]{Dimov2008}
\bibinfo{author}{\bibfnamefont{I.}~\bibnamefont{Dimov}},
  \bibinfo{author}{\bibfnamefont{P.}~\bibnamefont{Goswami}},
  \bibinfo{author}{\bibfnamefont{X.}~\bibnamefont{Jia}}, \bibnamefont{and}
  \bibinfo{author}{\bibfnamefont{S.}~\bibnamefont{Chakravarty}},
  \bibinfo{journal}{Phys. Rev. B} \textbf{\bibinfo{volume}{78}},
  \bibinfo{pages}{134529} (\bibinfo{year}{2008}).

\bibitem[{\citenamefont{Kurosawa et~al.}(2010)\citenamefont{Kurosawa, Yoneyama,
  Takano, Hagiwara, Inoue, Hagiwara, Kurusu, Takeyama, Momono, Oda, and
  Ido}}]{Kurosawa2001}
\bibinfo{author}{\bibfnamefont{T.}~\bibnamefont{Kurosawa}},
  \bibinfo{author}{\bibfnamefont{T.}~\bibnamefont{Yoneyama}},
  \bibinfo{author}{\bibfnamefont{Y.}~\bibnamefont{Takano}},
  \bibinfo{author}{\bibfnamefont{M.}~\bibnamefont{Hagiwara}},
  \bibinfo{author}{\bibfnamefont{R.}~\bibnamefont{Inoue}},
  \bibinfo{author}{\bibfnamefont{N.}~\bibnamefont{Hagiwara}},
  \bibinfo{author}{\bibfnamefont{K.}~\bibnamefont{Kurusu}},
  \bibinfo{author}{\bibfnamefont{K.}~\bibnamefont{Takeyama}},
  \bibinfo{author}{\bibfnamefont{N.}~\bibnamefont{Momono}},
  \bibinfo{author}{\bibfnamefont{M.}~\bibnamefont{Oda}}, \bibnamefont{and}
  \bibinfo{author}{\bibfnamefont{M.}~\bibnamefont{Ido}},
  \bibinfo{journal}{Phys. Rev. B} \textbf{\bibinfo{volume}{81}},
  \bibinfo{pages}{094519} (\bibinfo{year}{2010}).

\bibitem[{\citenamefont{Doiron-Leyraud
  et~al.}(2007)\citenamefont{Doiron-Leyraud, Proust, LeBoeuf, Levallois,
  Bonnemaison, Liang, Bonn, Hardy, and Taillefer}}]{Doiron2007}
\bibinfo{author}{\bibfnamefont{N.}~\bibnamefont{Doiron-Leyraud}},
  \bibinfo{author}{\bibfnamefont{C.}~\bibnamefont{Proust}},
  \bibinfo{author}{\bibfnamefont{D.}~\bibnamefont{LeBoeuf}},
  \bibinfo{author}{\bibfnamefont{J.}~\bibnamefont{Levallois}},
  \bibinfo{author}{\bibfnamefont{J.-B.} \bibnamefont{Bonnemaison}},
  \bibinfo{author}{\bibfnamefont{R.}~\bibnamefont{Liang}},
  \bibinfo{author}{\bibfnamefont{D.~A.} \bibnamefont{Bonn}},
  \bibinfo{author}{\bibfnamefont{W.~N.} \bibnamefont{Hardy}}, \bibnamefont{and}
  \bibinfo{author}{\bibfnamefont{L.}~\bibnamefont{Taillefer}},
  \bibinfo{journal}{Nature (London)} \textbf{\bibinfo{volume}{447}},
  \bibinfo{pages}{565} (\bibinfo{year}{2007}).

\bibitem[{\citenamefont{Sebastian et~al.}(2012)\citenamefont{Sebastian,
  Harrison, and Lonzarich}}]{Sebastian2012}
\bibinfo{author}{\bibfnamefont{S.~E.} \bibnamefont{Sebastian}},
  \bibinfo{author}{\bibfnamefont{N.}~\bibnamefont{Harrison}}, \bibnamefont{and}
  \bibinfo{author}{\bibfnamefont{G.}~\bibnamefont{Lonzarich}},
  \bibinfo{journal}{Rep. Prog. Phys.} \textbf{\bibinfo{volume}{75}},
  \bibinfo{pages}{102501} (\bibinfo{year}{2012}).

\bibitem[{\citenamefont{Chakravarty
  et~al.}(2001{\natexlab{b}})\citenamefont{Chakravarty, Kee, and
  Nayak}}]{Chakravarty2001b}
\bibinfo{author}{\bibfnamefont{S.}~\bibnamefont{Chakravarty}},
  \bibinfo{author}{\bibfnamefont{H.-Y.} \bibnamefont{Kee}}, \bibnamefont{and}
  \bibinfo{author}{\bibfnamefont{C.}~\bibnamefont{Nayak}},
  \bibinfo{journal}{Int. J. Mod. Phys. B} \textbf{\bibinfo{volume}{15}},
  \bibinfo{pages}{2901} (\bibinfo{year}{2001}{\natexlab{b}}).

\bibitem[{\citenamefont{Mook et~al.}(2001)\citenamefont{Mook, Dai, and
  Dogan}}]{Mook2001}
\bibinfo{author}{\bibfnamefont{H.~A.} \bibnamefont{Mook}},
  \bibinfo{author}{\bibfnamefont{P.}~\bibnamefont{Dai}}, \bibnamefont{and}
  \bibinfo{author}{\bibfnamefont{F.}~\bibnamefont{Dogan}},
  \bibinfo{journal}{Phys. Rev. B} \textbf{\bibinfo{volume}{64}},
  \bibinfo{pages}{012502} (\bibinfo{year}{2001}).

\bibitem[{\citenamefont{Stock et~al.}(2002)\citenamefont{Stock, Buyers, Tun,
  Liang, Peets, Bonn, Hardy, and Taillefer}}]{Stock2002}
\bibinfo{author}{\bibfnamefont{C.}~\bibnamefont{Stock}},
  \bibinfo{author}{\bibfnamefont{W.~J.~L.} \bibnamefont{Buyers}},
  \bibinfo{author}{\bibfnamefont{Z.}~\bibnamefont{Tun}},
  \bibinfo{author}{\bibfnamefont{R.}~\bibnamefont{Liang}},
  \bibinfo{author}{\bibfnamefont{D.}~\bibnamefont{Peets}},
  \bibinfo{author}{\bibfnamefont{D.}~\bibnamefont{Bonn}},
  \bibinfo{author}{\bibfnamefont{W.~N.} \bibnamefont{Hardy}}, \bibnamefont{and}
  \bibinfo{author}{\bibfnamefont{L.}~\bibnamefont{Taillefer}},
  \bibinfo{journal}{Phys. Rev. B} \textbf{\bibinfo{volume}{66}},
  \bibinfo{pages}{024505} (\bibinfo{year}{2002}).

\bibitem[{\citenamefont{H\"ucker et~al.}(2014)\citenamefont{H\"ucker,
  Christensen, Holmes, Blackburn, Forgan, Liang, Bonn, Hardy, Gutowski,
  Zimmermann, Hayden, and Chang}}]{Huecker2014}
\bibinfo{author}{\bibfnamefont{M.}~\bibnamefont{H\"ucker}},
  \bibinfo{author}{\bibfnamefont{N.~B.} \bibnamefont{Christensen}},
  \bibinfo{author}{\bibfnamefont{A.~T.} \bibnamefont{Holmes}},
  \bibinfo{author}{\bibfnamefont{E.}~\bibnamefont{Blackburn}},
  \bibinfo{author}{\bibfnamefont{E.~M.} \bibnamefont{Forgan}},
  \bibinfo{author}{\bibfnamefont{R.}~\bibnamefont{Liang}},
  \bibinfo{author}{\bibfnamefont{D.~A.} \bibnamefont{Bonn}},
  \bibinfo{author}{\bibfnamefont{W.~N.} \bibnamefont{Hardy}},
  \bibinfo{author}{\bibfnamefont{O.}~\bibnamefont{Gutowski}},
  \bibinfo{author}{\bibfnamefont{M.~v.} \bibnamefont{Zimmermann}},
  \bibinfo{author}{\bibfnamefont{S.~M.} \bibnamefont{Hayden}},
  \bibnamefont{and} \bibinfo{author}{\bibfnamefont{J.}~\bibnamefont{Chang}},
  \bibinfo{journal}{Phys. Rev. B} \textbf{\bibinfo{volume}{90}},
  \bibinfo{pages}{054514} (\bibinfo{year}{2014}).

\bibitem[{\citenamefont{Blanco-Canosa et~al.}(2014)\citenamefont{Blanco-Canosa,
  Frano, Schierle, Porras, Loew, Minola, Bluschke, Weschke, Keimer, and
  Le~Tacon}}]{Blanco2014}
\bibinfo{author}{\bibfnamefont{S.}~\bibnamefont{Blanco-Canosa}},
  \bibinfo{author}{\bibfnamefont{A.}~\bibnamefont{Frano}},
  \bibinfo{author}{\bibfnamefont{E.}~\bibnamefont{Schierle}},
  \bibinfo{author}{\bibfnamefont{J.}~\bibnamefont{Porras}},
  \bibinfo{author}{\bibfnamefont{T.}~\bibnamefont{Loew}},
  \bibinfo{author}{\bibfnamefont{M.}~\bibnamefont{Minola}},
  \bibinfo{author}{\bibfnamefont{M.}~\bibnamefont{Bluschke}},
  \bibinfo{author}{\bibfnamefont{E.}~\bibnamefont{Weschke}},
  \bibinfo{author}{\bibfnamefont{B.}~\bibnamefont{Keimer}}, \bibnamefont{and}
  \bibinfo{author}{\bibfnamefont{M.}~\bibnamefont{Le~Tacon}},
  \bibinfo{journal}{Phys. Rev. B} \textbf{\bibinfo{volume}{90}},
  \bibinfo{pages}{054513} (\bibinfo{year}{2014}).

\bibitem[{\citenamefont{Nie et~al.}(2014)\citenamefont{Nie, Tarjus, and
  Kivelson}}]{Nie:2014jj}
\bibinfo{author}{\bibfnamefont{L.}~\bibnamefont{Nie}},
  \bibinfo{author}{\bibfnamefont{G.}~\bibnamefont{Tarjus}}, \bibnamefont{and}
  \bibinfo{author}{\bibfnamefont{S.~A.} \bibnamefont{Kivelson}},
  \bibinfo{journal}{Proc.\ Nat.\ Acad.\ Sci.} \textbf{\bibinfo{volume}{111}},
  \bibinfo{pages}{7980} (\bibinfo{year}{2014}).

\end{thebibliography}
\bibliographystyle{apsrev_upto25authors}
\appendix


\section{Current operators}
\label{s:jops}
The current operator is conventionally defined as
\begin{equation}
	\hat J_{ij} = -it_{ij}(\hat c^\dagger_i \hat c^{}_j - \hat c^\dagger_j \hat c^{}_i ).
\end{equation}
At this point, however, it is important to take into account that we are working with a Hamiltonian which is gauge transformed according to Equations \reff{e:gtks1}-\reff{e:gtks4}.
In the transformed operator basis, the current operators in real space take the following form:
\begin{eqnarray}
J_{i d,j x}&=&t_{i d,j x}( \hat c^\dagger_{i d} \hat c_{j x} 
                + \hat c^\dagger_{j x} \hat c_{i d} ) \\
J_{i x,j d}&=&-t_{i x,j d}( \hat c^\dagger_{i d} \hat c_{j x} 
                + \hat c^\dagger_{j x} \hat c_{i d} ) \\
J_{i d,j y}&=&t_{i d,j y}( \hat c^\dagger_{i d} \hat c_{j y} 
                + \hat c^\dagger_{j y} \hat c_{i d} ) \\
J_{i y,j d}&=&-t_{i y,j d}( \hat c^\dagger_{i d} \hat c_{j y} 
                + \hat c^\dagger_{j y} \hat c_{i d} ) \\
J_{i x,j y}&=&-it_{i x,j y}( \hat c^\dagger_{i x} \hat c_{j y} 
                - \hat c^\dagger_{j y} \hat c_{i x} ) \\
J_{i y,j x}&=&-it_{i y,j x}( \hat c^\dagger_{i y} \hat c_{j x} 
                - \hat c^\dagger_{j x} \hat c_{i y} ). 
\end{eqnarray}


\section{Estimating the resulting magnetic moment}
\label{s:estim}
  
Here, we perform a simple estimate of the magnetic moment associated with the $\pi$LC loop currents.

We start by expressing the currents in absolute units.
The unit of the current operators is $e[t_{ij}]/[\hbar]=C(eV)/(eV\cdot s)=C/s$. 
For $z_{pd} = 0.04\tpd$, 
the current along $p$-$d$ bonds is 	$I_{pd}=z_{pd}\times(e/\hbar)=0.04\tpd\times(2.4\times10^{-4})A=9.6\tpd \mu A$, where $\tpd$ is measured in eV.  Band structure
calculations suggest $\tpd \approx 1.6$ eV, yielding $I_{pd} = 16$ $\mu$A; experimental bandwidths, however, are typically a factor of 3 smaller than predicted by band structure calculations suggesting $I_{pd} \sim 5$ $\mu$A.  This is comparable to other estimates of loop current amplitudes: 
within a cluster calculation of multi-orbital $t$-$J$ model, the upper bound of $\theta_{II}$-like loop currents was previously estimated to be between $5\mu A$ and $15\mu A$ for different parameter sets \cite{Thomale2008};
similarly, in the single band DDW studies, the staggered loop currents were estimated to be $\sim 7\mu A$ by assuming a DDW gap value of $\sim 0.03$eV\cite{Chakravarty2001}.

Next, we calculate the magnetic moments of loop currents using the formula $M=I\eta$ where $I$ is the current, and $\eta$ is the area enclosed by the loop.
This is indeed a crude calculation, however we believe this should yield a qualitatively correct number. 
The total magnetic moment in a given plaquette has two contrubutions: $I_b\eta_b$ and $I_g\eta_g$ due to two independent current loops (shown in black and green in Fig.~1).
We set $I_b=5\mu A$, and since it circulates around the whole plaquette, $\eta_b=a^2$.
As given by the eigenvectors of $\chi^J(q)$, $I_g\approx I_{pd}/3$, and it circulates around an area of $\eta_g=a^2/2$.
Thus, the magnetic moment is calculated as
\begin{eqnarray}
    M 
	&=& I_b\eta_b + I_g\eta_g \\
	&=& I_{pd}\times a^2 +I_{pd}/3\times a^2/2 \\
    &=& 7I_{pd}a^2/6 \\
    &=& 7(5 \mu A) (3.85 \times 10^{-10}m)^2/6 \\
    &\approx& 0.09\mu_B.
\end{eqnarray}
However, since $I_{pd}$ is shared by the neighbouring plaquettes, the effective moment for individual plaquettes might be reduced to $\sim0.05\mu_B$.
\end{document}